\documentclass[letter]{aa} 
\usepackage{txfonts}
\usepackage{graphicx, graphics}
\usepackage{CJK}
\usepackage{bbm}
\usepackage{amsmath, amssymb, bm}
\usepackage[nodayofweek]{datetime}
\input{prepcitationformat.txt}
\usepackage{txfonts}
\usepackage{changes}

\usepackage{mathtools}

\DeclarePairedDelimiterX\braket[2]{\langle}{\rangle}{#1 \delimsize\vert #2}
\newcommand{\Qphi}{$\mathcal{Q}_\phi$}

\usepackage[switch]{lineno} 

\usepackage{tikzsymbols}
\usepackage{tikz}    
\newcommand{\ExternalLinkSmall}{\!%
    \tikz[x=0.9ex, y=0.9ex, baseline=-0.9ex, blue]{%
        \begin{scope}[x=0.7ex, y=0.7ex]
            \clip (-0.1,-0.1) 
                --++ (-0, 1.2) 
                --++ (0.6, 0) 
                --++ (0, -0.6) 
                --++ (0.6, 0) 
                --++ (0, -1);
            \path[draw, 
                line width = 1, 
                rounded corners=1] 
                (0,0) rectangle (1,1);
        \end{scope}
        \path[draw, line width = 1] (0.5, 0.5) 
            -- (1, 1);
        \path[draw, line width = 1] (0.6, 1) 
            -- (1, 1) -- (1, 0.6);
        }\!\!\!
    }

\definecolor{lime}{HTML}{A6CE39}
\DeclareRobustCommand{\orcidicon}{%
    \raisebox{-3pt}{\begin{tikzpicture}
    \filldraw [lime, yshift=-2pt] (0, 0) circle [radius=0.16]
    node[white] {\raisebox{1pt}{\hspace{0.5pt}\fontfamily{qag}\selectfont\tiny i\scalebox{0.8}{D}}};
    \end{tikzpicture}}
    \hspace{-2.5mm}
    \vspace{-0.25pt}
}

\newdateformat{monthyearday}{
  \THEYEAR\ \monthname[\THEMONTH] \twodigit{\THEDAY}}
\newdateformat{daymonthyear}{%
  \twodigit{\THEDAY}\ \monthname[\THEMONTH] \THEYEAR}

\newcommand{\orcidauthor}[2]{#2\href{http://orcid.org/#1}{\orcidicon}}
\DeclareRobustCommand{\ion}[2]{%
\relax\ifmmode
\ifx\testbx\f@series
{\mathbf{#1\,\mathsc{#2}}}\else
{\mathrm{#1\,\mathsc{#2}}}\fi
\else\textup{#1\,{\mdseries\textsc{#2}}}%
\fi}

\newcommand{\programESO}[1]{#1~\href{http://archive.eso.org/wdb/wdb/eso/sched_rep_arc/query?progid=#1}{\ExternalLinkSmall}}

\titlerunning{V1247~Ori Spiral Motion}
\authorrunning{Ren et al.}

\begin{document}
\begin{CJK*}{UTF8}{gbsn}
\title{A Companion in V1247~Ori Supported by Spiral Arm Pattern Motion\thanks{Based on observations performed with VLT/SPHERE under program ID 0102.C-0778 and 111.24GG.}}

\author{
\orcidauthor{0000-0003-1698-9696}{Bin B. Ren (任彬)\thanks{Marie Sk\l odowska-Curie Fellow}$^,$\thanks{To whom correspondence should be addressed.}}\inst{\ref{inst-oca}, \ref{inst-uga}}
\and
\orcidauthor{0000-0002-6318-0104}{Chen Xie (谢晨)$^{\star\star\star}$}\inst{\ref{inst-jhu}, \ref{inst-lam}}
\and
\orcidauthor{0000-0002-7695-7605}{Myriam Benisty}\inst{\ref{inst-oca}, \ref{inst-uga}} 
\and
\orcidauthor{0000-0001-9290-7846}{Ruobing Dong  (董若冰)}\inst{\ref{inst-uvic}}
\and
\orcidauthor{0000-0001-7258-770X}{Jaehan Bae}\inst{\ref{inst-ufl}}
\and
\orcidauthor{0000-0002-5823-3072}{Tomas Stolker}\inst{\ref{leiden}}
\and
\orcidauthor{0000-0003-1520-8405}{Rob G. van Holstein}\inst{\ref{inst-eso}}
\and
\orcidauthor{0000-0002-1783-8817}{John H. Debes}\inst{\ref{inst-stsci}}
\and
\orcidauthor{0000-0002-4266-0643}{Antonio Garufi}\inst{\ref{inst-inaf}}
\and
\orcidauthor{0000-0002-4438-1971}{Christian Ginski}\inst{\ref{inst-galway}}
\and
\orcidauthor{0000-0001-6017-8773}{Stefan Kraus}\inst{\ref{inst-exter}}
}

\institute{
Universit\'{e} C\^{o}te d'Azur, Observatoire de la C\^{o}te d'Azur, CNRS, Laboratoire Lagrange, Bd de l'Observatoire, CS 34229, F-06304 Nice cedex 4, France; 
\email{\url{bin.ren@oca.eu}} \label{inst-oca}
\and
Universit\'{e} Grenoble Alpes, CNRS, Institut de Plan\'{e}tologie et d'Astrophysique (IPAG), F-38000 Grenoble, France \label{inst-uga}
\and
Department of Physics and Astronomy, The Johns Hopkins University, 3701 San Martin Drive, Baltimore, MD 21218, USA; \email{\url{chen.xie@lam.fr}} \label{inst-jhu}
\and
Aix Marseille Univ, CNRS, CNES, LAM, Marseille, France\label{inst-lam}
\and
Department of Physics \& Astronomy, University of Victoria, Victoria, BC, V8P 5C2, Canada \label{inst-uvic}
\and
Department of Astronomy, University of Florida, Gainesville, FL 32611, USA \label{inst-ufl}
\and
Leiden Observatory, Leiden University, Niels Bohrweg 2, 2333 CA Leiden, The Netherlands \label{leiden}
\and
European Southern Observatory, Alonso de C\'ordova 3107, Vitacura Casilla 19001, Santiago, Chile \label{inst-eso}
\and
AURA, for ESA, Space Telescope Science Institute, 3700 San Martin Dr., Baltimore, MD 21218, USA \label{inst-stsci}
\and
INAF, Osservatorio Astrofisico di Arcetri, Largo Enrico Fermi 5, I-50125 Firenze, Italy \label{inst-inaf}
\and
School of Natural Sciences, University of Galway, University Road, H91 TK33 Galway, Ireland \label{inst-galway}
\and
Astrophysics Group, Department of Physics \& Astronomy, University of Exeter, Stocker Road, Exeter EX4 4QL, UK \label{inst-exter}
}

\date{Received 29 September 2023 / Revised 10 November 2023 / Accepted 6 December 2023}

\abstract
{While there have been nearly two dozen of spiral arms detected from planet-forming disks in near-infrared scattered light, none of their substellar drivers have been confirmed. By observing spiral systems in at least two epochs spanning multiple years, and measuring the motion of the spirals, we can distinguish the cause of the spirals, and locate the orbits of the driving planets if they trigger the spirals. Upon a recent validation of this approach using the co-motion between a stellar companion and a spiral, we obtained a second epoch observation for the spiral system in the disk of V1247 Ori in the $H$-band polarized scattered light  using VLT/SPHERE/IRDIS. Combining our observations with archival IRDIS data, we established a $4.8$~yr timeline to constrain the V1247~Ori spiral motion. We obtained a pattern speed of $0\fdg40\pm0\fdg10$~yr$^{-1}$ for the north-east spiral. This corresponds to an orbital period of $900\pm220$~yr, and thus the semi-major axis of the hidden planetary driver is $118\pm20$~au for a $2.0\pm0.1~M_\sun$ central star. The location agrees with the gap in ALMA dust continuum observations, providing joint support for the existence of a companion driving the scattered-light spirals while carving a millimeter gap. With an angular separation of $0\farcs29\pm0\farcs05$, this hidden companion is an ideal target for \textit{JWST} imaging.}


\maketitle

\section{Introduction}
Over the course of the past decade of direct imaging attempts for self-luminous planets using state-of-the-art high-contrast imaging instruments \citep[e.g.,][]{nielsen20, vigan21}, only a handful new discoveries are confirmed \citep[e.g.,][]{macintosh15, chauvin17, keppler18, haffert19}. The low detection rate
called for optimized and targeted observing strategies in more recent studies, including young systems \citep[e.g.,][]{bohn21}, stellar radial velocity \citep[e.g.,][]{lagrange19, nowak20} and astrometric acceleration \citep[e.g.,][]{currie23, franson23, derosa23, mesa23} evidences, and morphological deviation of circumstellar disks \citep[e.g.,][]{bae19, haffert19, wang20}.

\begin{figure*}[htb!]
	\centering
	\includegraphics[width=0.7\textwidth]{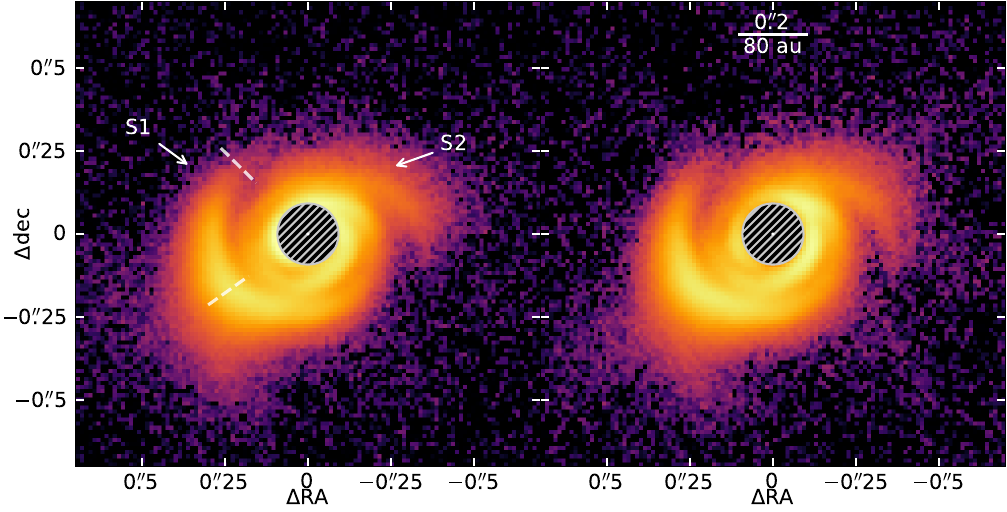}
    \caption{Two epochs \Qphi\ images of V1247~Ori with SPHERE/IRDIS in $H$-band on 2018 November 16 (left), and on 2023 September 20 (right). The image color bars are in log scale. The dashed lines are the boundaries for motion measurement in Fig.~\ref{fig2}.
    \\ (The data used to create this figure are available.)}
    \label{fig1}
\end{figure*}

Targeted imaging based on deviations from stellar signals (e.g., radial velocity, astrometry) have produced fruitful results. Nevertheless, monitoring these signals may take up to multiple decades to investigate planetary existence \citep[e.g.,][]{lagrange19, currie23}, and they rely on instruments whose stability and sensitivity could have improved significantly over time. Meanwhile, substructures engraved on circumstellar disks by planets have suggested planetary potential existence \citep[e.g.,][]{dong15planet, dong16, bae16, zhang18, long22}. Among planet-induced structures, spiral arms could provide best systems for targeted high-contrast imaging search, since their planetary drivers tend to be more massive \citep[e.g.,][]{bae18} and thus more luminous \citep[e.g.,][]{spiegel12}, which are more likely to be accessed by current generation of high-contrast imaging instruments.

The pattern motion of a spiral arm, when driven by a companion, follows the orbital motion of its driver \citep[e.g.,][]{dong15planet, bae18spiral}. By tracing the motion of spirals, we can constrain the orbit of a hidden driver \citep{ren18}, and distinguish the motion between leading mechanisms \citep[i.e., companion-driven and gravity-instatbility-induction:][]{ren20}. By showing the first dynamic evidence of the co-motion between a stellar companion and a spiral in the HD~100453 system, \citet{xie23} confirmed the theory of companion-driven spirals \citep[e.g.,][]{dong15planet, dong16, bae16}. This consolidates that spiral motion monitoring is a major science case for future extremely large telescopes \citep{maire21}. After the first light of current high-contrast imagers in ${\sim}2015$, we can now re-image the spirals with identical instrument across an ${\approx}5$~yr separation \citep[e.g.,][]{ren20, xie23}, and thus we should expect a rapid expansion of such studies \citep{benisty23}. Once a spiral system is confirmed to be driven by hidden planetary companions, we can conduct targeted imaging efforts for their potential existence \citep[e.g.,][]{wagner19, wagner23, boccaletti21}.

Among known spiral arm hosts \citep[e.g.,][]{dong18, shuai22}, \object{V1247~Ori} is an A5III or F0V star \citep[e.g.,][]{kharchenko01, vieira03} located at $401\pm3$~pc \citep{GaiaDR3}. With an estimated age of $7.0\pm0.3$~Myr \citep{garufi18}, it hosts a pair of nearly symmetric spiral arms in scattered light \citep[e.g., $H$-band: ][]{shuai22} and a gap in ALMA submillimeter/millimeter observations \citep[e.g., Band~7: ][]{kraus17}. Based on the dust continuum morphology, \citet{kraus17} proposed a planet at ${\sim}0\farcs3$, or $120$~au with updated distances from \textit{Gaia}~DR3, from the star could carve the ALMA gap. While it is possible that the V1247~Ori spirals could be triggered by a recent stellar flyby (i.e., ${\sim}10^4$~yr in time, and ${\sim}10\times$ disk radius in separation), using the location and proper motion measurements from \textit{Gaia}~DR3, \citet{shuai22} identified no flyby events for V1247~Ori. In addition, the spirals might be triggered by gravitational instability, yet the disk-to-star mass ratio should be $q \gtrsim0.25$ in \citet{dong15gi}, which contradicts the $q\approx0.04$ from \citet{yu19}. These combined evidences make V1247~Ori a probable system that hosts a planetary companion that drives the spirals in scattered light. Observationally, by re-imaging V1247~Ori in scattered light at another epoch, we will be able to test whether the spiral motion is consistent with being driven by a planet in the ALMA dust continuum gap.

To observationally constrain the motion pattern for existing spiral systems, SAFFRON (Spiral Arm Formation from mOtion aNalysis; ESO programs 111.24GG and 112.25B7) is a dedicated survey which acquires additional epoch(s) of existing spirals with a total allocation of $54.5$~hours from VLT/SPHERE/IRDIS. For these systems, we can use spiral motion to distinguish the motion mechanisms between companion-driven and gravitational-instability-induction \citep[][]{ren20}, predict the orbits of hidden drivers \citep[e.g.,][]{ren18, xie21}, and confirm the connection between known companion(s) and spiral(s) \citep{xie23}. Through SAFFRON observations, we aim to constrain the pattern motion for all the spiral systems in near-infrared scattered light \citep[e.g.,][]{shuai22}. What is more, if the motions are consistent with being driven by companions, we will be able to map the orbital distribution of these hidden planets. In this Letter, we report the first result from the SAFFRON survey for V1247~Ori.

\section{Observation \& Data Reduction}
Using the Very Large Telescope (VLT), we observed V1247~Ori using the infrared dual-band imager and spectrograph (IRDIS; \citealp{dohlen08}) on SPHERE \citep{beuzit19} with the dual-beam polarimetric imaging (DPI; \citealp{deboer20, irdap2}) mode at $H$-band (1.625~$\mu$m) on UT 2023 September 20 under ESO program 111.24GG in SAFFRON (PI: B.~Ren). The observation complements the one on 2018 November 16 under ESO program 0102.C-0778 (PI: S.~Kraus) to enable precise spiral pattern motion measurement. 

To ensure similar data quality for spiral pattern motion measurement, we minimized the observational difference as follows. Specifically, we adopted the same $32$ s exposure in each integration. There are $9$ polarimetric cycles through $Q^+$, $Q^-$, $U^+$, and $U^-$ with $3$ integrations per exposure in 2018, and $15$ polarimetric cycles with $1$ integration per exposure in 2023. We totaled $3456$ s on-target time in 2018 and $1920$ s in 2023. At the end of the observations, there are $128$ s and $32$ s total exposure, respectively, for sky background removal. 

We reduced the two observed IRDIS DPI data sets using the {\tt IRDAP} pipeline \citep{irdap1, irdap2}. In the reduction, we also adopted the parallactic angles calculated using the \citet{pynpoint} pipeline for SPHERE. To enable precise spiral location and motion measurement in further analysis, we additionally stretched the preprocessed cubes by $1.006$ along the column direction of the detector to correct for the anamorphism from optical distortions in SPHERE \citep[e.g.,][]{schmid18}. Among the post-processed data products from {\tt IRDAP}, we used the star-polarization-subtracted local Stokes \Qphi\ map, which traces the distribution of dust particles on the surfaces of disks \citep[e.g.,][]{monnier19}, for analysis. We present the \Qphi\ images from the two epochs and annotate the spiral arms in Fig.~\ref{fig1}.

To reduce the impact of stellar light illumination as a function of stellocentric radius, namely the incident flux reduces as distance squared, we should use the surface density distribution images for motion analysis \citep[e.g.,][]{xie23}. Specifically, we scaled the surface brightness distribution for the two \Qphi\ images of V1247~Ori. We adopted an inclination of $30\fdg0\pm1\fdg0$ from face-on, and a position angle of $-64\fdg6\pm0\fdg4$ for the major axis of the disk from north to east from ALMA analysis \citep[][]{kraus17}. To correct for disk flaring effects, we assumed that the disk scale height, $h$, follows $h/r = 0.05$, where $r$ is radial separation \citep{kraus17}. We then computed the mid-plane stellocentric distances, $r$, for all image pixels, and multiplied the value at each pixel by $r^2$ to reveal the distribution of the scatterers which typically trace surface morphology and roughly trace dust surface density.

\section{Analysis}

To measure spiral location, we analyzed the $r^2$-scaled \Qphi\ images as follows. First, we deprojected the disk images to face-on views. Second, we determined the local maxima of the spiral arm in polar coordinates: we performed three Gaussian profile fits with an extra constant, in a total of ten free parameters \citep[e.g.,][]{ren20, xie23}. The extra constant takes into account the overall disk brightness offset at a specific location. Specifically, given that there could exist multiple spirals at a specific azimuthal angle, we adopted three Gaussian profiles for the two wrapped arm components along the radial direction (e.g., the ${\sim}2$ o'clock region in Fig.~\ref{fig1}). Third, in each azimuthal angle, we performed fittings to obtain the locations of the spiral arms with a $1^\circ$ angular step, we present the locations for a selected region of the northeast spiral S1 in Fig.~\ref{fig2}. For measurements on the entirety of the two spirals, see Figs.~\ref{fig-a2} and \ref{fig-a3} in Appendix~\ref{sec-app}. However, given that the two spirals are visually merged interior to ${\approx}125$~au due to the finite resolution of VLT/SPHERE, the motion measurements for the entirety of the two spirals are likely biased, and thus we only focus on the exterior region of S1 marked in Fig.~\ref{fig1} in this study.

\begin{figure}[htb!]
	\centering
	\includegraphics[width=0.51\textwidth]{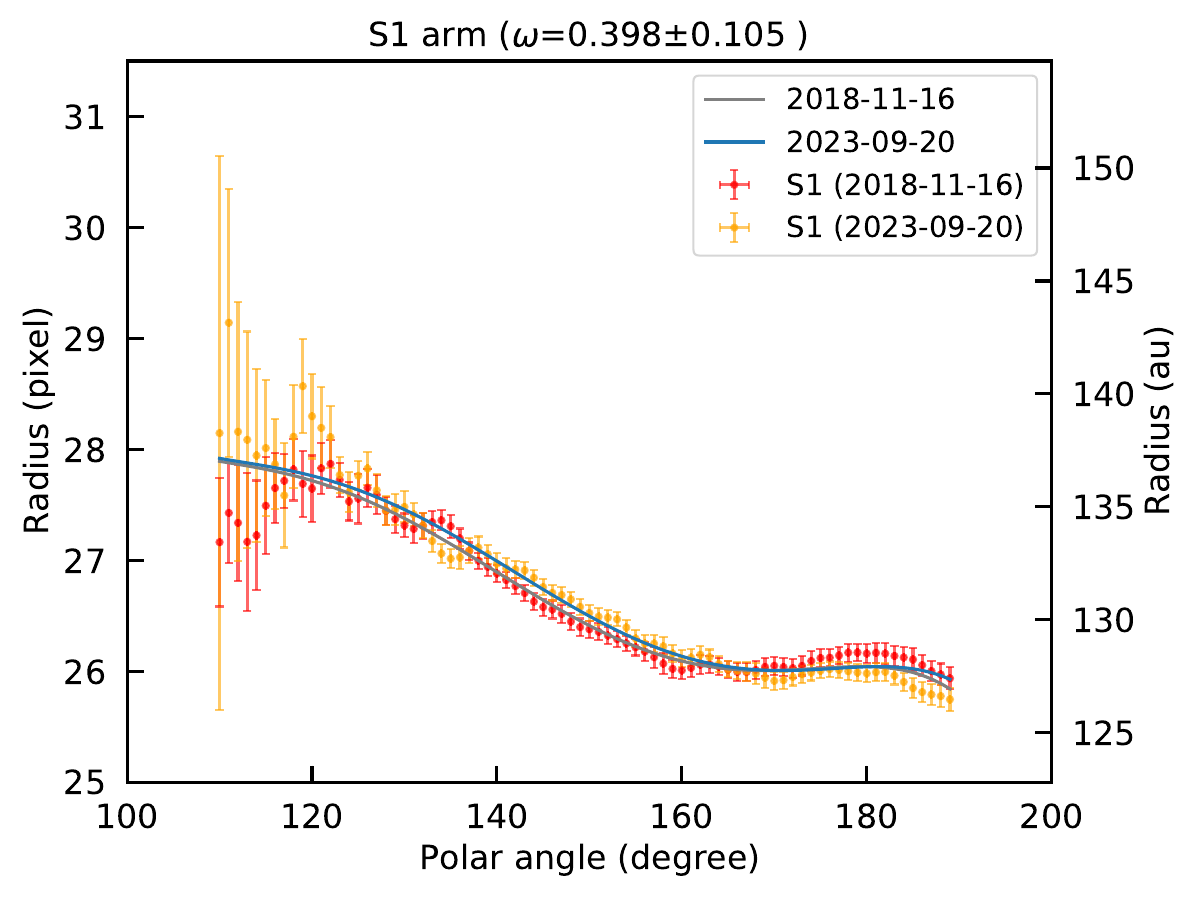}
    \caption{Peak locations of spiral arm S1 between 110$^\circ$ and 190$^\circ$ in polar coordinates in disk plane. The solid curves represent the best-fit model spiral for the peak locations (dot points) between 2018 and 2023 ($t=4.84$~yr), assuming the companion-driven scenario. The derived angular velocity of spiral pattern motion is $0\fdg40\pm0\fdg10~$yr$^{-1}$ counterclockwise. }
    \label{fig2}
\end{figure}

To simultaneously derive the spiral pattern motion and its morphological parameters, we fitted $p$-degree polynomials with a dummy variable to a given spiral arm in two epochs as described in \cite{ren20}. The Schwarz information criterion, which penalizes the excessive use of parameters, is minimized at $p=5$ and thus we adopted the corresponding results for motion analysis. Using the dummy variable, we can simultaneously fit an identical but offset polynomial to a spiral at both epochs. The large extent of the spiral arm S1 enables us to explore motion measurements for different parts of the spiral, see Fig.~\ref{fig2} for a specific area of S1, and Fig.~\ref{fig-a2} for the entirety of S1. 

Spiral location and pattern speed measurement can be influenced by multiple factors for V1247~Ori. On the one hand, with an ${\approx}30^\circ$ inclination \citep[e.g.,][]{kraus17, bohn22}, the spiral morphology of V1247~Ori could be distorted \citep[limiting inclination of ${\approx}20^\circ$:][]{dong16spiralprojection}. On the other hand, we witnessed global shadowing effects in the north-east regions in Fig.~\ref{fig1}, which could be cast by inner disks \citep[e.g.,][]{debes17, debes23}. Due to the non-negligible inclination of the system (${\approx}30^\circ$) and the given angular resolution of SPHERE ($51$~mas in $H$-band for an $8.0$~m effective pupil), the location and motion measurement for the entirety of S1 and S2 in Figs.~\ref{fig-a2} and \ref{fig-a3} could thus be biased. Specifically, spiral morphology can be altered when the disk is not seen face-on \citep[${\gtrsim}20^\circ$ inclination:][]{dong16spiralprojection}, and if spirals are driven by an identical but eccentric companion \citep[e.g.,][]{calcino20}, \citet{xie23} showed that spirals could have different apparent motion.

\begin{figure}[hbt!]
	\centering
	\includegraphics[width=0.4\textwidth]{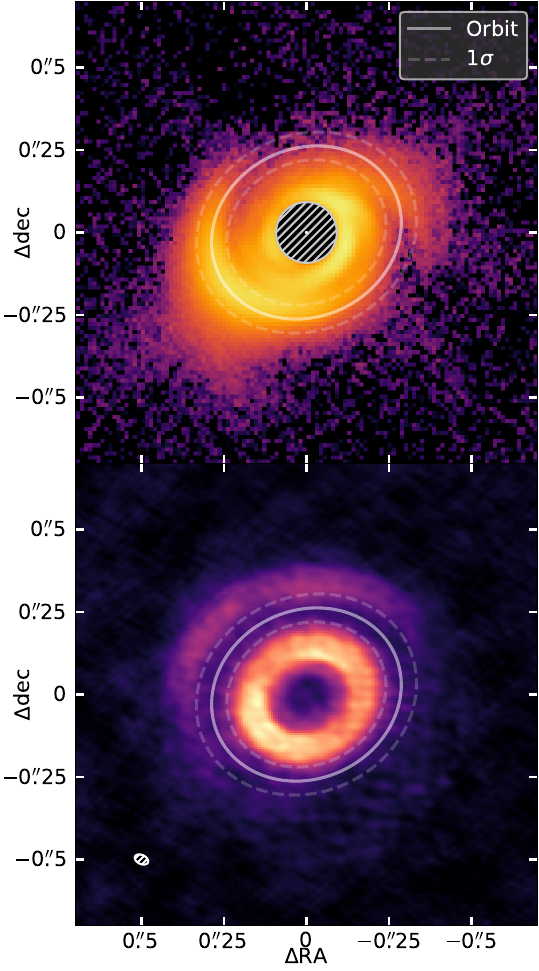}
    \caption{Driver orbit from S1 spiral pattern motion, overlaid on SPHERE $H$-band \Qphi\ (top; UT 2023 September 20) and ALMA Band 7 continuum (bottom; UT 2017 August 09) observations. For a $2.0\pm0.1~M_\sun$ central star, the $0\fdg40\pm0\fdg10$~yr$^{-1}$ counterclockwise S1 spiral motion corresponds to a driver at $118\pm20$~au, or $0\farcs29\pm0\farcs05$. Subject to motion measurement uncertainties here, the derived orbit is consistent with the ${\approx}120$~au ALMA gap-opener proposed in \citet{kraus17} within $1\sigma$. }
    \label{fig-orbit}
\end{figure}

To quantify the spiral motion, we only focused on a specific region of the S1 spiral in Fig.~\ref{fig2} for motion measurement. As has been investigated in \citet{xie23}, we should measure the spiral location and motion for a disk surface that is approximately face-on through a combination of disk inclination and disk flaring. For this purpose, we focused on a spiral region of S1 that has a stellocentric radius of $125$--$140$~au, which spans an azimuthal coverage of ${\approx}80^\circ$ from the disk midplane. In this region, the inclination and flaring of the spiral make the disk surface closer to face-on view, and we measured its spiral motion of S1 to be\footnote{The derived uncertainties presented in this Letter are $1\sigma$ unless otherwise specified.}


\begin{equation}\label{eq-motion}
\omega = 0\fdg40\pm0\fdg10~{\rm yr}^{-1},
\end{equation} counterclockwise in Fig.~\ref{fig2}. We obtained the uncertainty in Eq.~\eqref{eq-motion} using
$\sigma =\sqrt{(\sigma_{\text{fit} }t)^{2} +2\sigma^{2}_{\text{north} } +2\sigma^{2}_{\text{centering} }} \  t^{-1}$, 
where $\sigma_{\rm fit}$ and $\sigma_{\rm north}$ are uncertainties caused by the measurement of the spiral locations and true north uncertainty of SPHERE, respectively. The time span between two epochs is represented by $t$, which is $4.84$~yr. The selected S1 locations return a fitting uncertainty $\sigma_{\rm fit} = 0\fdg091$~yr$^{-1}$. We adopt the true north uncertainty of SPHERE to be $\sigma_{\text{north} } = 0\fdg08$ in all epochs \citep{maire16}. We also adopted an astrometric uncertainty of ${\approx}2$~mas \citep[][Fig.~7 therein]{zurlo14}, or $0.16$ IRDIS pixel and thus an offset of $\sigma_{\text{centering}}=0\fdg16$ in polar coordinates. The combined uncertainty in motion rate is thus $\sigma=0\fdg10$~yr$^{-1}$.  

Data sampling and disk parameters may affect the spiral motion measurements. Different data sampling (e.g., Nyquist sampling: 5$^\circ$~angular step) gives a consistent spiral motion of $0\fdg46\pm0\fdg18$~yr$^{-1}$ with a slightly larger total uncertainty. Correlated noise in the oversampling case (1$^\circ$~angular step) may lead to the underestimation of the total uncertainty. However, given the consistent motion measurements and the minor impact on the uncertainty, proper estimation of the correlated noise \citep[e.g.,][]{Czekala2015} in the oversampling case is outside the scope of this work. The impact from disk parameters (e.g., inclination, flaring, position angle) has been extensively analyzed (e.g., Sect.~4.5 of \citealp{ren20}, Sect.~4.2 of \citealp{xie21}, and Appendix~D of \citealp{xie23}), showing that the corresponding impact resides within the derived motion uncertainty. 

The S1 pattern motion in the selected region in Fig.~\ref{fig1} is $\omega = 0\fdg40\pm0\fdg10~{\rm yr}^{-1}$ in Eq.~\eqref{eq-motion}, which corresponds to a driver at $118\pm20$~au on a circular orbit from a central star with $2.0\pm0.1~M_\sun$ \citep{garufi18}. To illustrate its orbit, we obtained the ALMA Band 7 continuum observations of V1247~Ori from \citet{kraus17}. Together with the scattered light \Qphi\ image, we present the predicted driver orbit in Fig.~\ref{fig-orbit}.

Our measurement in Fig.~\ref{fig2} cannot confidently rule out local gravitational motion, with the local Keplerian motion at 125 au ($0\farcs3$) to 140~au ($0\farcs35$) to be $0\fdg349$~yr$^{-1}$ to $0\fdg296$~yr$^{-1}$, respectively.  However, the disk-to-star mass ratio of V1247~Ori is $q\approx0.04$ from \citet{yu19}, making it unlikely to trigger local gravitational instability for spiral motion. Were V1247~Ori spirals triggered by gravitational instability, the disk-to-star mass ratio should be $q \gtrsim0.25$ in \citet{dong15gi}, which is ${\sim}6$ times more massive than the current ratio and thus unlikely.

In comparison with the ALMA data, for a star with a mass of $2.0\pm0.1~M_\sun$, if the proposed gap-opening planet in \citet{kraus17} drives the spiral, the pattern motion of the spiral should be $0\fdg39\pm0\fdg01$~yr$^{-1}$. This is consistent with the pattern motion measurement in Eq.~\eqref{eq-motion} within $1\sigma$. With our motion measurement, we now can offer a dynamical evidence that there should exist a spiral-arm-driving planet in scattered light, and that driver simultaneously carves the millimeter gap in \citet{kraus17}.

\section{Conclusion}
The pattern motion of spirals can inform their formation mechanisms \citep[e.g.,][]{ren20}. If driven by a companion, then the motion of the spiral pattern traces the orbit of its driver \citep[e.g.,][]{ren18, safonov22}. With two epochs of polarized light observations spanning $4.84$~yr, we constrained the pattern motion for the S1 spiral in scattered light in the V1247~Ori system to be $0\fdg40\pm0\fdg10~$yr$^{-1}$ counterclockwise. For a $2.0\pm0.1M_\sun$ star, this translates to a semi-major axis of $118\pm20$~au, or $0\farcs29\pm0\farcs05$, for the hidden spiral-arm-driving companion.

The spiral-arm-driving companion in Fig.~\ref{fig-orbit} is not yet imaged for V1247~Ori in ground-based observatories \citep[e.g.,][]{ohta16, willson19, follette23, wallack23}. Meanwhile, based on misalignments between inner and outer disks \citep[e.g.,][]{bohn22, woelfer23}, even if there exists an inner companion that creates the inner cavity in ALMA and drive the spirals behind the coronagraph (${\sim}0\farcs1$) in Fig.~\ref{fig1}, we can rule out such a companion-spiral connection at ${>5}\sigma$ level from our motion measurements. 

Being a suspect that both drives spiral(s) in scattered light and carves a gap in millimeter continuum, this hidden companion is the first one that has two different observational evidences to support its existence. Due to the sensitivity limits with ground-based instruments for this driver (${\sim}80$ Jupiter mass, \citealp{ren23}), and given the demonstrated superior sensitivity of the \textit{JWST}/NIRCam coronagraphic mode in \citet{carter23}, together with the fact that planetary drivers for a pair of symmetric spirals should be of $5$--$10$ Jupiter mass \citep[i.e.,][]{dong15planet, bae16}, the located driver for V1247~Ori here serves an ideal target for \textit{JWST}/NIRCam coronagraphic imaging.

\begin{acknowledgements}
We thank the anonymous referee for their comments and suggestions that improved the quality of this Letter. We thank Andrew Winter and Jochen Stadler for discussions. Based on observations collected at the European Organisation for Astronomical Research in the Southern Hemisphere under ESO programmes \programESO{0102.C-0778} and \programESO{111.24GG}. This project has received funding from the European Union's Horizon Europe research and innovation programme under the Marie Sk\l odowska-Curie grant agreement No.~101103114. This project has received funding from the European Research Council (ERC) under the European Union's Horizon 2020 research and innovation programme (PROTOPLANETS, grant agreement No.~101002188). R.D.~acknowledges financial support provided by the Natural Sciences and Engineering Research Council of Canada through a Discovery Grant, as well as the Alfred P.~Sloan Foundation through a Sloan Research Fellowship. This paper makes use of the following ALMA data: ADS/JAO.ALMA\#2015.1.00986.S. ALMA is a partnership of ESO (representing its member states), NSF (USA) and NINS (Japan), together with NRC (Canada), MOST and ASIAA (Taiwan, ROC), and KASI (Republic of Korea), in cooperation with the Republic of Chile. The Joint ALMA Observatory is operated by ESO, AUI/NRAO and NAOJ. This work has made use of data from the European Space Agency (ESA) mission {\it Gaia} (\url{https://www.cosmos.esa.int/gaia}), processed by the {\it Gaia} Data Processing and Analysis Consortium (DPAC, \url{https://www.cosmos.esa.int/web/gaia/dpac/consortium}).\end{acknowledgements}

\bibliography{refs}

\begin{thebibliography}{}
\expandafter\ifx\csname natexlab\endcsname\relax\def\natexlab#1{#1}\fi
\providecommand{\url}[1]{\href{#1}{#1}}
\providecommand{\dodoi}[1]{}
\providecommand{\doarXiv}[1]{\href{https://arxiv.org/abs/#1}{\nolinkurl{https://arxiv.org/abs/#1}}}

\bibitem[{{Bae} {et~al.}(2018){Bae}, {Pinilla}, \& {Birnstiel}}]{bae18}
{Bae}, J., {Pinilla}, P., \& {Birnstiel}, T. 2018,
  \href{http://dx.doi.org/10.3847/2041-8213/aadd51}{\color{magenta}\apjl},
  \href{https://ui.adsabs.harvard.edu/abs/2018ApJ...864L..26B}{\color{blue}864},
  \href{https://ui.adsabs.harvard.edu/abs/2018ApJ...864L..26B}{\color{blue}L26}

\bibitem[{{Bae} \& {Zhu}(2018)}]{bae18spiral}
{Bae}, J., \& {Zhu}, Z. 2018,
  \href{http://dx.doi.org/10.3847/1538-4357/aabf93}{\color{magenta}\apj},
  \href{https://ui.adsabs.harvard.edu/abs/2018ApJ...859..119B}{\color{blue}859},
  \href{https://ui.adsabs.harvard.edu/abs/2018ApJ...859..119B}{\color{blue}119}

\bibitem[{{Bae} {et~al.}(2016){Bae}, {Zhu}, \& {Hartmann}}]{bae16}
{Bae}, J., {Zhu}, Z., \& {Hartmann}, L. 2016,
  \href{http://dx.doi.org/10.3847/0004-637X/819/2/134}{\color{magenta}\apj},
  \href{https://ui.adsabs.harvard.edu/abs/2016ApJ...819..134B}{\color{blue}819},
  \href{https://ui.adsabs.harvard.edu/abs/2016ApJ...819..134B}{\color{blue}134}

\bibitem[{{Bae} {et~al.}(2019){Bae}, {Zhu}, {Baruteau}, {Benisty}, {Dullemond},
  {Facchini}, {Isella}, {Keppler}, {P{\'e}rez}, \& {Teague}}]{bae19}
{Bae}, J., {Zhu}, Z., {Baruteau}, C., {et~al.} 2019,
  \href{http://dx.doi.org/10.3847/2041-8213/ab46b0}{\color{magenta}\apjl},
  \href{https://ui.adsabs.harvard.edu/abs/2019ApJ...884L..41B}{\color{blue}884},
  \href{https://ui.adsabs.harvard.edu/abs/2019ApJ...884L..41B}{\color{blue}L41}

\bibitem[{{Benisty} {et~al.}(2023){Benisty}, {Dominik}, {Follette}, {Garufi},
  {Ginski}, {Hashimoto}, {Keppler}, {Kley}, \& {Monnier}}]{benisty23}
{Benisty}, M., {Dominik}, C., {Follette}, K., {et~al.} 2023,
  \href{http://dx.doi.org/10.48550/arXiv.2203.09991}{\color{magenta}Astronomical
  Society of the Pacific Conference Series},
  \href{https://ui.adsabs.harvard.edu/abs/2023ASPC..534..605B}{\color{blue}534},
  \href{https://ui.adsabs.harvard.edu/abs/2023ASPC..534..605B}{\color{blue}605}

\bibitem[{{Beuzit} {et~al.}(2019){Beuzit}, {Vigan}, {Mouillet}, {Dohlen},
  {Gratton}, {Boccaletti}, {Sauvage}, {Schmid}, {Langlois}, {Petit},
  {Baruffolo}, {Feldt}, {Milli}, {Wahhaj}, {Abe}, {Anselmi}, {Antichi},
  {Barette}, {Baudrand}, {Baudoz}, {Bazzon}, {Bernardi}, {Blanchard}, {Brast},
  {Bruno}, {Buey}, {Carbillet}, {Carle}, {Cascone}, {Chapron}, {Charton},
  {Chauvin}, {Claudi}, {Costille}, {De Caprio}, {de Boer}, {Delboulb{\'e}},
  {Desidera}, {Dominik}, {Downing}, {Dupuis}, {Fabron}, {Fantinel}, {Farisato},
  {Feautrier}, {Fedrigo}, {Fusco}, {Gigan}, {Ginski}, {Girard}, {Giro},
  {Gisler}, {Gluck}, {Gry}, {Henning}, {Hubin}, {Hugot}, {Incorvaia}, {Jaquet},
  {Kasper}, {Lagadec}, {Lagrange}, {Le Coroller}, {Le Mignant}, {Le Ruyet},
  {Lessio}, {Lizon}, {Llored}, {Lundin}, {Madec}, {Magnard}, {Marteaud},
  {Martinez}, {Maurel}, {M{\'e}nard}, {Mesa}, {M{\"o}ller-Nilsson}, {Moulin},
  {Moutou}, {Orign{\'e}}, {Parisot}, {Pavlov}, {Perret}, {Pragt}, {Puget},
  {Rabou}, {Ramos}, {Reess}, {Rigal}, {Rochat}, {Roelfsema}, {Rousset}, {Roux},
  {Saisse}, {Salasnich}, {Santambrogio}, {Scuderi}, {Segransan}, {Sevin},
  {Siebenmorgen}, {Soenke}, {Stadler}, {Suarez}, {Tiph{\`e}ne}, {Turatto},
  {Udry}, {Vakili}, {Waters}, {Weber}, {Wildi}, {Zins}, \& {Zurlo}}]{beuzit19}
{Beuzit}, J.~L., {Vigan}, A., {Mouillet}, D., {et~al.} 2019,
  \href{http://dx.doi.org/10.1051/0004-6361/201935251}{\color{magenta}\aap},
  \href{https://ui.adsabs.harvard.edu/abs/2019A&A...631A.155B}{\color{blue}631},
  \href{https://ui.adsabs.harvard.edu/abs/2019A&A...631A.155B}{\color{blue}A155}

\bibitem[{{Boccaletti} {et~al.}(2021){Boccaletti}, {Pantin}, {M{\'e}nard},
  {Galicher}, {Langlois}, {Benisty}, {Gratton}, {Chauvin}, {Ginski},
  {Lagrange}, {Zurlo}, {Biller}, {Bonavita}, {Bonnefoy}, {Brown-Sevilla},
  {Cantalloube}, {Desidera}, {D'Orazi}, {Feldt}, {Hagelberg}, {Lazzoni},
  {Mesa}, {Meyer}, {Perrot}, {Vigan}, {Sauvage}, {Ramos}, {Rousset}, \&
  {Magnard}}]{boccaletti21}
{Boccaletti}, A., {Pantin}, E., {M{\'e}nard}, F., {et~al.} 2021,
  \href{http://dx.doi.org/10.1051/0004-6361/202141177}{\color{magenta}\aap},
  \href{https://ui.adsabs.harvard.edu/abs/2021A&A...652L...8B}{\color{blue}652},
  \href{https://ui.adsabs.harvard.edu/abs/2021A&A...652L...8B}{\color{blue}L8}

\bibitem[{{Bohn} {et~al.}(2021){Bohn}, {Ginski}, {Kenworthy}, {Mamajek},
  {Pecaut}, {Mugrauer}, {Vogt}, {Adam}, {Meshkat}, {Reggiani}, \&
  {Snik}}]{bohn21}
{Bohn}, A.~J., {Ginski}, C., {Kenworthy}, M.~A., {et~al.} 2021,
  \href{http://dx.doi.org/10.1051/0004-6361/202140508}{\color{magenta}\aap},
  \href{https://ui.adsabs.harvard.edu/abs/2021A&A...648A..73B}{\color{blue}648},
  \href{https://ui.adsabs.harvard.edu/abs/2021A&A...648A..73B}{\color{blue}A73}

\bibitem[{{Bohn} {et~al.}(2022){Bohn}, {Benisty}, {Perraut}, {van der Marel},
  {W{\"o}lfer}, {van Dishoeck}, {Facchini}, {Manara}, {Teague}, {Francis},
  {Berger}, {Garcia-Lopez}, {Ginski}, {Henning}, {Kenworthy}, {Kraus},
  {M{\'e}nard}, {M{\'e}rand}, \& {P{\'e}rez}}]{bohn22}
{Bohn}, A.~J., {Benisty}, M., {Perraut}, K., {et~al.} 2022,
  \href{http://dx.doi.org/10.1051/0004-6361/202142070}{\color{magenta}\aap},
  \href{https://ui.adsabs.harvard.edu/abs/2022A&A...658A.183B}{\color{blue}658},
  \href{https://ui.adsabs.harvard.edu/abs/2022A&A...658A.183B}{\color{blue}A183}

\bibitem[{{Calcino} {et~al.}(2020){Calcino}, {Christiaens}, {Price}, {Pinte},
  {Davis}, {van der Marel}, \& {Cuello}}]{calcino20}
{Calcino}, J., {Christiaens}, V., {Price}, D.~J., {et~al.} 2020,
  \href{http://dx.doi.org/10.1093/mnras/staa2468}{\color{magenta}\mnras},
  \href{https://ui.adsabs.harvard.edu/abs/2020MNRAS.498..639C}{\color{blue}498},
  \href{https://ui.adsabs.harvard.edu/abs/2020MNRAS.498..639C}{\color{blue}639}

\bibitem[{{Carter} {et~al.}(2023){Carter}, {Hinkley}, {Kammerer}, {Skemer},
  {Biller}, {Leisenring}, {Millar-Blanchaer}, {Petrus}, {Stone}, {Ward-Duong},
  {Wang}, {Girard}, {Hines}, {Perrin}, {Pueyo}, {Balmer}, {Bonavita},
  {Bonnefoy}, {Chauvin}, {Choquet}, {Christiaens}, {Danielski}, {Kennedy},
  {Matthews}, {Miles}, {Patapis}, {Ray}, {Rickman}, {Sallum}, {Stapelfeldt},
  {Whiteford}, {Zhou}, {Absil}, {Boccaletti}, {Booth}, {Bowler}, {Chen},
  {Currie}, {Fortney}, {Grady}, {Greebaum}, {Henning}, {Hoch}, {Janson},
  {Kalas}, {Kenworthy}, {Kervella}, {Kraus}, {Lagage}, {Liu}, {Macintosh},
  {Marino}, {Marley}, {Marois}, {Matthews}, {Mawet}, {McElwain}, {Metchev},
  {Meyer}, {Molliere}, {Moran}, {Morley}, {Mukherjee}, {Pantin}, {Quirrenbach},
  {Rebollido}, {Ren}, {Schneider}, {Vasist}, {Worthen}, {Wyatt},
  {Briesemeister}, {Bryan}, {Calissendorff}, {Cantalloube}, {Cugno}, {De
  Furio}, {Dupuy}, {Factor}, {Faherty}, {Fitzgerald}, {Franson}, {Gonzales},
  {Hood}, {Howe}, {Kuzuhara}, {Lagrange}, {Lawson}, {Lazzoni}, {Lew}, {Liu},
  {Llop-Sayson}, {Lloyd}, {Martinez}, {Mazoyer}, {Palma-Bifani}, {Quanz},
  {Redai}, {Samland}, {Schlieder}, {Tamura}, {Tan}, {Uyama}, {Vigan}, {Vos},
  {Wagner}, {Wolff}, {Ygouf}, {Zhang}, {Zhang}, \& {Zhang}}]{carter23}
{Carter}, A.~L., {Hinkley}, S., {Kammerer}, J., {et~al.} 2023,
  \href{http://dx.doi.org/10.3847/2041-8213/acd93e}{\color{magenta}\apjl},
  \href{https://ui.adsabs.harvard.edu/abs/2023ApJ...951L..20C}{\color{blue}951},
  \href{https://ui.adsabs.harvard.edu/abs/2023ApJ...951L..20C}{\color{blue}L20}

\bibitem[{{Chauvin} {et~al.}(2017){Chauvin}, {Desidera}, {Lagrange}, {Vigan},
  {Gratton}, {Langlois}, {Bonnefoy}, {Beuzit}, {Feldt}, {Mouillet}, {Meyer},
  {Cheetham}, {Biller}, {Boccaletti}, {D'Orazi}, {Galicher}, {Hagelberg},
  {Maire}, {Mesa}, {Olofsson}, {Samland}, {Schmidt}, {Sissa}, {Bonavita},
  {Charnay}, {Cudel}, {Daemgen}, {Delorme}, {Janin-Potiron}, {Janson},
  {Keppler}, {Le Coroller}, {Ligi}, {Marleau}, {Messina}, {Molli{\`e}re},
  {Mordasini}, {M{\"u}ller}, {Peretti}, {Perrot}, {Rodet}, {Rouan}, {Zurlo},
  {Dominik}, {Henning}, {Menard}, {Schmid}, {Turatto}, {Udry}, {Vakili}, {Abe},
  {Antichi}, {Baruffolo}, {Baudoz}, {Baudrand}, {Blanchard}, {Bazzon}, {Buey},
  {Carbillet}, {Carle}, {Charton}, {Cascone}, {Claudi}, {Costille}, {Deboulbe},
  {De Caprio}, {Dohlen}, {Fantinel}, {Feautrier}, {Fusco}, {Gigan}, {Giro},
  {Gisler}, {Gluck}, {Hubin}, {Hugot}, {Jaquet}, {Kasper}, {Madec}, {Magnard},
  {Martinez}, {Maurel}, {Le Mignant}, {M{\"o}ller-Nilsson}, {Llored}, {Moulin},
  {Orign{\'e}}, {Pavlov}, {Perret}, {Petit}, {Pragt}, {Puget}, {Rabou},
  {Ramos}, {Rigal}, {Rochat}, {Roelfsema}, {Rousset}, {Roux}, {Salasnich},
  {Sauvage}, {Sevin}, {Soenke}, {Stadler}, {Suarez}, {Weber}, {Wildi},
  {Antoniucci}, {Augereau}, {Baudino}, {Brandner}, {Engler}, {Girard}, {Gry},
  {Kral}, {Kopytova}, {Lagadec}, {Milli}, {Moutou}, {Schlieder},
  {Szul{\'a}gyi}, {Thalmann}, \& {Wahhaj}}]{chauvin17}
{Chauvin}, G., {Desidera}, S., {Lagrange}, A.~M., {et~al.} 2017,
  \href{http://dx.doi.org/10.1051/0004-6361/201731152}{\color{magenta}\aap},
  \href{https://ui.adsabs.harvard.edu/abs/2017A&A...605L...9C}{\color{blue}605},
  \href{https://ui.adsabs.harvard.edu/abs/2017A&A...605L...9C}{\color{blue}L9}

\bibitem[{{Currie} {et~al.}(2023){Currie}, {Brandt}, {Brandt}, {Lacy},
  {Burrows}, {Guyon}, {Tamura}, {Liu}, {Sagynbayeva}, {Tobin}, {Chilcote},
  {Groff}, {Marois}, {Thompson}, {Murphy}, {Kuzuhara}, {Lawson}, {Lozi}, {Deo},
  {Vievard}, {Skaf}, {Uyama}, {Jovanovic}, {Martinache}, {Kasdin}, {Kudo},
  {McElwain}, {Janson}, {Wisniewski}, {Hodapp}, {Nishikawa}, {He{\l}miniak},
  {Kwon}, \& {Hayashi}}]{currie23}
{Currie}, T., {Brandt}, G.~M., {Brandt}, T.~D., {et~al.} 2023,
  \href{http://dx.doi.org/10.1126/science.abo6192}{\color{magenta}Science},
  \href{https://ui.adsabs.harvard.edu/abs/2023Sci...380..198C}{\color{blue}380},
  \href{https://ui.adsabs.harvard.edu/abs/2023Sci...380..198C}{\color{blue}198}

\bibitem[{{Czekala} {et~al.}(2015){Czekala}, {Andrews}, {Mandel}, {Hogg}, \&
  {Green}}]{Czekala2015}
{Czekala}, I., {Andrews}, S.~M., {Mandel}, K.~S., {et~al.} 2015,
  \href{http://dx.doi.org/10.1088/0004-637X/812/2/128}{\color{magenta}\apj},
  \href{https://ui.adsabs.harvard.edu/abs/2015ApJ...812..128C}{\color{blue}812},
  \href{https://ui.adsabs.harvard.edu/abs/2015ApJ...812..128C}{\color{blue}128}

\bibitem[{{de Boer} {et~al.}(2020){de Boer}, {Langlois}, {van Holstein},
  {Girard}, {Mouillet}, {Vigan}, {Dohlen}, {Snik}, {Keller}, {Ginski}, {Stam},
  {Milli}, {Wahhaj}, {Kasper}, {Schmid}, {Rabou}, {Gluck}, {Hugot}, {Perret},
  {Martinez}, {Weber}, {Pragt}, {Sauvage}, {Boccaletti}, {Le Coroller},
  {Dominik}, {Henning}, {Lagadec}, {M{\'e}nard}, {Turatto}, {Udry}, {Chauvin},
  {Feldt}, \& {Beuzit}}]{deboer20}
{de Boer}, J., {Langlois}, M., {van Holstein}, R.~G., {et~al.} 2020,
  \href{http://dx.doi.org/10.1051/0004-6361/201834989}{\color{magenta}\aap},
  \href{https://ui.adsabs.harvard.edu/abs/2020A&A...633A..63D}{\color{blue}633},
  \href{https://ui.adsabs.harvard.edu/abs/2020A&A...633A..63D}{\color{blue}A63}

\bibitem[{{De Rosa} {et~al.}(2023){De Rosa}, {Nielsen}, {Wahhaj}, {Ruffio},
  {Kalas}, {Peck}, {Hirsch}, \& {Roberson}}]{derosa23}
{De Rosa}, R.~J., {Nielsen}, E.~L., {Wahhaj}, Z., {et~al.} 2023,
  \href{http://dx.doi.org/10.1051/0004-6361/202345877}{\color{magenta}\aap},
  \href{https://ui.adsabs.harvard.edu/abs/2023A&A...672A..94D}{\color{blue}672},
  \href{https://ui.adsabs.harvard.edu/abs/2023A&A...672A..94D}{\color{blue}A94}

\bibitem[{{Debes} {et~al.}(2023){Debes}, {Nealon}, {Alexander}, {Weinberger},
  {Wolff}, {Hines}, {Kastner}, {Jang-Condell}, {Pinte}, {Plavchan}, \&
  {Pueyo}}]{debes23}
{Debes}, J., {Nealon}, R., {Alexander}, R., {et~al.} 2023,
  \href{http://dx.doi.org/10.3847/1538-4357/acbdf1}{\color{magenta}\apj},
  \href{https://ui.adsabs.harvard.edu/abs/2023ApJ...948...36D}{\color{blue}948},
  \href{https://ui.adsabs.harvard.edu/abs/2023ApJ...948...36D}{\color{blue}36}

\bibitem[{{Debes} {et~al.}(2017){Debes}, {Poteet}, {Jang-Condell}, {Gaspar},
  {Hines}, {Kastner}, {Pueyo}, {Rapson}, {Roberge}, {Schneider}, \&
  {Weinberger}}]{debes17}
{Debes}, J.~H., {Poteet}, C.~A., {Jang-Condell}, H., {et~al.} 2017,
  \href{http://dx.doi.org/10.3847/1538-4357/835/2/205}{\color{magenta}\apj},
  \href{https://ui.adsabs.harvard.edu/abs/2017ApJ...835..205D}{\color{blue}835},
  \href{https://ui.adsabs.harvard.edu/abs/2017ApJ...835..205D}{\color{blue}205}

\bibitem[{{Dohlen} {et~al.}(2008){Dohlen}, {Langlois}, {Saisse}, {Hill},
  {Origne}, {Jacquet}, {Fabron}, {Blanc}, {Llored}, {Carle}, {Moutou}, {Vigan},
  {Boccaletti}, {Carbillet}, {Mouillet}, \& {Beuzit}}]{dohlen08}
{Dohlen}, K., {Langlois}, M., {Saisse}, M., {et~al.} 2008,
  \href{http://dx.doi.org/10.1117/12.789786}{\color{magenta}Proc.~SPIE},
  \href{https://ui.adsabs.harvard.edu/abs/2008SPIE.7014E..3LD}{\color{blue}7014},
  \href{https://ui.adsabs.harvard.edu/abs/2008SPIE.7014E..3LD}{\color{blue}70143L}

\bibitem[{{Dong} {et~al.}(2016{\natexlab{a}}){Dong}, {Fung}, \&
  {Chiang}}]{dong16spiralprojection}
{Dong}, R., {Fung}, J., \& {Chiang}, E. 2016{\natexlab{a}},
  \href{http://dx.doi.org/10.3847/0004-637X/826/1/75}{\color{magenta}\apj},
  \href{https://ui.adsabs.harvard.edu/abs/2016ApJ...826...75D}{\color{blue}826},
  \href{https://ui.adsabs.harvard.edu/abs/2016ApJ...826...75D}{\color{blue}75}

\bibitem[{{Dong} {et~al.}(2015{\natexlab{a}}){Dong}, {Hall}, {Rice}, \&
  {Chiang}}]{dong15gi}
{Dong}, R., {Hall}, C., {Rice}, K., \& {Chiang}, E. 2015{\natexlab{a}},
  \href{http://dx.doi.org/10.1088/2041-8205/812/2/L32}{\color{magenta}\apjl},
  \href{https://ui.adsabs.harvard.edu/abs/2015ApJ...812L..32D}{\color{blue}812},
  \href{https://ui.adsabs.harvard.edu/abs/2015ApJ...812L..32D}{\color{blue}L32}

\bibitem[{{Dong} {et~al.}(2018){Dong}, {Najita}, \& {Brittain}}]{dong18}
{Dong}, R., {Najita}, J.~R., \& {Brittain}, S. 2018,
  \href{http://dx.doi.org/10.3847/1538-4357/aaccfc}{\color{magenta}\apj},
  \href{https://ui.adsabs.harvard.edu/abs/2018ApJ...862..103D}{\color{blue}862},
  \href{https://ui.adsabs.harvard.edu/abs/2018ApJ...862..103D}{\color{blue}103}

\bibitem[{{Dong} {et~al.}(2016{\natexlab{b}}){Dong}, {Zhu}, {Fung}, {Rafikov},
  {Chiang}, \& {Wagner}}]{dong16}
{Dong}, R., {Zhu}, Z., {Fung}, J., {et~al.} 2016{\natexlab{b}},
  \href{http://dx.doi.org/10.3847/2041-8205/816/1/L12}{\color{magenta}\apjl},
  \href{https://ui.adsabs.harvard.edu/abs/2016ApJ...816L..12D}{\color{blue}816},
  \href{https://ui.adsabs.harvard.edu/abs/2016ApJ...816L..12D}{\color{blue}L12}

\bibitem[{{Dong} {et~al.}(2015{\natexlab{b}}){Dong}, {Zhu}, {Rafikov}, \&
  {Stone}}]{dong15planet}
{Dong}, R., {Zhu}, Z., {Rafikov}, R.~R., \& {Stone}, J.~M. 2015{\natexlab{b}},
  \href{http://dx.doi.org/10.1088/2041-8205/809/1/L5}{\color{magenta}\apjl},
  \href{https://ui.adsabs.harvard.edu/abs/2015ApJ...809L...5D}{\color{blue}809},
  \href{https://ui.adsabs.harvard.edu/abs/2015ApJ...809L...5D}{\color{blue}L5}

\bibitem[{{Follette} {et~al.}(2023){Follette}, {Close}, {Males}, {Ward-Duong},
  {Balmer}, {Redai}, {Morales}, {Sarosi}, {Dacus}, {De Rosa}, {Garcia Toro},
  {Leonard}, {Macintosh}, {Morzinski}, {Mullen}, {Palmo}, {Saitoti}, {Spiro},
  {Treiber}, {Wagner}, {Wang}, {Wang}, {Watson}, \& {Weinberger}}]{follette23}
{Follette}, K.~B., {Close}, L.~M., {Males}, J.~R., {et~al.} 2023,
  \href{http://dx.doi.org/10.3847/1538-3881/acc183}{\color{magenta}\aj},
  \href{https://ui.adsabs.harvard.edu/abs/2023AJ....165..225F}{\color{blue}165},
  \href{https://ui.adsabs.harvard.edu/abs/2023AJ....165..225F}{\color{blue}225}

\bibitem[{{Franson} {et~al.}(2023){Franson}, {Bowler}, {Zhou}, {Pearce},
  {Bardalez Gagliuffi}, {Biddle}, {Brandt}, {Crepp}, {Dupuy}, {Faherty},
  {Jensen-Clem}, {Morgan}, {Sanghi}, {Theissen}, {Tran}, \& {Wolf}}]{franson23}
{Franson}, K., {Bowler}, B.~P., {Zhou}, Y., {et~al.} 2023,
  \href{http://dx.doi.org/10.3847/2041-8213/acd6f6}{\color{magenta}\apjl},
  \href{https://ui.adsabs.harvard.edu/abs/2023ApJ...950L..19F}{\color{blue}950},
  \href{https://ui.adsabs.harvard.edu/abs/2023ApJ...950L..19F}{\color{blue}L19}

\bibitem[{{\textit{Gaia} Collaboration} {et~al.}(2023){\textit{Gaia}
  Collaboration}, {Vallenari}, {Brown}, {Prusti}, {de Bruijne}, {Arenou},
  {Babusiaux}, {Biermann}, {Creevey}, {Ducourant}, \& et~al.}]{GaiaDR3}
{\textit{Gaia} Collaboration}, {Vallenari}, A., {Brown}, A.~G.~A., {et~al.}
  2023,
  \href{http://dx.doi.org/10.1051/0004-6361/202243940}{\color{magenta}\aap},
  \href{https://ui.adsabs.harvard.edu/abs/2023A&A...674A...1G}{\color{blue}674},
  \href{https://ui.adsabs.harvard.edu/abs/2023A&A...674A...1G}{\color{blue}A1}

\bibitem[{{Garufi} {et~al.}(2018){Garufi}, {Benisty}, {Pinilla}, {Tazzari},
  {Dominik}, {Ginski}, {Henning}, {Kral}, {Langlois}, {M{\'e}nard}, {Stolker},
  {Szulagyi}, {Villenave}, \& {van der Plas}}]{garufi18}
{Garufi}, A., {Benisty}, M., {Pinilla}, P., {et~al.} 2018,
  \href{http://dx.doi.org/10.1051/0004-6361/201833872}{\color{magenta}\aap},
  \href{https://ui.adsabs.harvard.edu/abs/2018A&A...620A..94G}{\color{blue}620},
  \href{https://ui.adsabs.harvard.edu/abs/2018A&A...620A..94G}{\color{blue}A94}

\bibitem[{{Haffert} {et~al.}(2019){Haffert}, {Bohn}, {de Boer}, {Snellen},
  {Brinchmann}, {Girard}, {Keller}, \& {Bacon}}]{haffert19}
{Haffert}, S.~Y., {Bohn}, A.~J., {de Boer}, J., {et~al.} 2019,
  \href{http://dx.doi.org/10.1038/s41550-019-0780-5}{\color{magenta}NatAs},
  \href{https://ui.adsabs.harvard.edu/abs/2019NatAs...3..749H}{\color{blue}3},
  \href{https://ui.adsabs.harvard.edu/abs/2019NatAs...3..749H}{\color{blue}749}

\bibitem[{{Keppler} {et~al.}(2018){Keppler}, {Benisty}, {M{\"u}ller},
  {Henning}, {van Boekel}, {Cantalloube}, {Ginski}, {van Holstein}, {Maire},
  {Pohl}, {Samland}, {Avenhaus}, {Baudino}, {Boccaletti}, {de Boer},
  {Bonnefoy}, {Chauvin}, {Desidera}, {Langlois}, {Lazzoni}, {Marleau},
  {Mordasini}, {Pawellek}, {Stolker}, {Vigan}, {Zurlo}, {Birnstiel},
  {Brandner}, {Feldt}, {Flock}, {Girard}, {Gratton}, {Hagelberg}, {Isella},
  {Janson}, {Juhasz}, {Kemmer}, {Kral}, {Lagrange}, {Launhardt}, {Matter},
  {M{\'e}nard}, {Milli}, {Molli{\`e}re}, {Olofsson}, {P{\'e}rez}, {Pinilla},
  {Pinte}, {Quanz}, {Schmidt}, {Udry}, {Wahhaj}, {Williams}, {Buenzli},
  {Cudel}, {Dominik}, {Galicher}, {Kasper}, {Lannier}, {Mesa}, {Mouillet},
  {Peretti}, {Perrot}, {Salter}, {Sissa}, {Wildi}, {Abe}, {Antichi},
  {Augereau}, {Baruffolo}, {Baudoz}, {Bazzon}, {Beuzit}, {Blanchard}, {Brems},
  {Buey}, {De Caprio}, {Carbillet}, {Carle}, {Cascone}, {Cheetham}, {Claudi},
  {Costille}, {Delboulb{\'e}}, {Dohlen}, {Fantinel}, {Feautrier}, {Fusco},
  {Giro}, {Gluck}, {Gry}, {Hubin}, {Hugot}, {Jaquet}, {Le Mignant}, {Llored},
  {Madec}, {Magnard}, {Martinez}, {Maurel}, {Meyer}, {M{\"o}ller-Nilsson},
  {Moulin}, {Mugnier}, {Orign{\'e}}, {Pavlov}, {Perret}, {Petit}, {Pragt},
  {Puget}, {Rabou}, {Ramos}, {Rigal}, {Rochat}, {Roelfsema}, {Rousset}, {Roux},
  {Salasnich}, {Sauvage}, {Sevin}, {Soenke}, {Stadler}, {Suarez}, {Turatto}, \&
  {Weber}}]{keppler18}
{Keppler}, M., {Benisty}, M., {M{\"u}ller}, A., {et~al.} 2018,
  \href{http://dx.doi.org/10.1051/0004-6361/201832957}{\color{magenta}\aap},
  \href{https://ui.adsabs.harvard.edu/abs/2018A&A...617A..44K}{\color{blue}617},
  \href{https://ui.adsabs.harvard.edu/abs/2018A&A...617A..44K}{\color{blue}A44}

\bibitem[{{Kharchenko}(2001)}]{kharchenko01}
{Kharchenko}, N.~V. 2001, Kinematika i Fizika Nebesnykh Tel,
  \href{https://ui.adsabs.harvard.edu/abs/2001KFNT...17..409K}{\color{blue}17},
  \href{https://ui.adsabs.harvard.edu/abs/2001KFNT...17..409K}{\color{blue}409}

\bibitem[{{Kraus} {et~al.}(2017){Kraus}, {Kreplin}, {Fukugawa}, {Muto},
  {Sitko}, {Young}, {Bate}, {Grady}, {Harries}, {Monnier}, {Willson}, \&
  {Wisniewski}}]{kraus17}
{Kraus}, S., {Kreplin}, A., {Fukugawa}, M., {et~al.} 2017,
  \href{http://dx.doi.org/10.3847/2041-8213/aa8edc}{\color{magenta}\apjl},
  \href{https://ui.adsabs.harvard.edu/abs/2017ApJ...848L..11K}{\color{blue}848},
  \href{https://ui.adsabs.harvard.edu/abs/2017ApJ...848L..11K}{\color{blue}L11}

\bibitem[{{Lagrange} {et~al.}(2019){Lagrange}, {Meunier}, {Rubini}, {Keppler},
  {Galland}, {Chapellier}, {Michel}, {Balona}, {Beust}, {Guillot}, {Grandjean},
  {Borgniet}, {M{\'e}karnia}, {Wilson}, {Kiefer}, {Bonnefoy}, {Lillo-Box},
  {Pantoja}, {Jones}, {Iglesias}, {Rodet}, {Diaz}, {Zapata}, {Abe}, \&
  {Schmider}}]{lagrange19}
{Lagrange}, A.~M., {Meunier}, N., {Rubini}, P., {et~al.} 2019,
  \href{http://dx.doi.org/10.1038/s41550-019-0857-1}{\color{magenta}NatAs},
  \href{https://ui.adsabs.harvard.edu/abs/2019NatAs...3.1135L}{\color{blue}3},
  \href{https://ui.adsabs.harvard.edu/abs/2019NatAs...3.1135L}{\color{blue}1135}

\bibitem[{{Long} {et~al.}(2022){Long}, {Andrews}, {Zhang}, {Qi}, {Benisty},
  {Facchini}, {Isella}, {Wilner}, {Bae}, {Huang}, {Loomis}, {{\"O}berg}, \&
  {Zhu}}]{long22}
{Long}, F., {Andrews}, S.~M., {Zhang}, S., {et~al.} 2022,
  \href{http://dx.doi.org/10.3847/2041-8213/ac8b10}{\color{magenta}\apjl},
  \href{https://ui.adsabs.harvard.edu/abs/2022ApJ...937L...1L}{\color{blue}937},
  \href{https://ui.adsabs.harvard.edu/abs/2022ApJ...937L...1L}{\color{blue}L1}

\bibitem[{{Macintosh} {et~al.}(2015){Macintosh}, {Graham}, {Barman}, {De Rosa},
  {Konopacky}, {Marley}, {Marois}, {Nielsen}, {Pueyo}, {Rajan}, {Rameau},
  {Saumon}, {Wang}, {Patience}, {Ammons}, {Arriaga}, {Artigau}, {Beckwith},
  {Brewster}, {Bruzzone}, {Bulger}, {Burningham}, {Burrows}, {Chen}, {Chiang},
  {Chilcote}, {Dawson}, {Dong}, {Doyon}, {Draper}, {Duch{\^e}ne}, {Esposito},
  {Fabrycky}, {Fitzgerald}, {Follette}, {Fortney}, {Gerard}, {Goodsell},
  {Greenbaum}, {Hibon}, {Hinkley}, {Cotten}, {Hung}, {Ingraham},
  {Johnson-Groh}, {Kalas}, {Lafreniere}, {Larkin}, {Lee}, {Line}, {Long},
  {Maire}, {Marchis}, {Matthews}, {Max}, {Metchev}, {Millar-Blanchaer},
  {Mittal}, {Morley}, {Morzinski}, {Murray-Clay}, {Oppenheimer}, {Palmer},
  {Patel}, {Perrin}, {Poyneer}, {Rafikov}, {Rantakyr{\"o}}, {Rice}, {Rojo},
  {Rudy}, {Ruffio}, {Ruiz}, {Sadakuni}, {Saddlemyer}, {Salama}, {Savransky},
  {Schneider}, {Sivaramakrishnan}, {Song}, {Soummer}, {Thomas}, {Vasisht},
  {Wallace}, {Ward-Duong}, {Wiktorowicz}, {Wolff}, \&
  {Zuckerman}}]{macintosh15}
{Macintosh}, B., {Graham}, J.~R., {Barman}, T., {et~al.} 2015,
  \href{http://dx.doi.org/10.1126/science.aac5891}{\color{magenta}Science},
  \href{https://ui.adsabs.harvard.edu/abs/2015Sci...350...64M}{\color{blue}350},
  \href{https://ui.adsabs.harvard.edu/abs/2015Sci...350...64M}{\color{blue}64}

\bibitem[{{Maire} {et~al.}(2021){Maire}, {Langlois}, {Delorme}, {Chauvin},
  {Gratton}, \& {Vigan}}]{maire21}
{Maire}, A.~L., {Langlois}, M., {Delorme}, P., {et~al.} 2021, SF2A-2021:
  Proceedings of the Annual meeting of the French Society of Astronomy and
  Astrophysics. Eds.: A. Siebert,
  \href{https://ui.adsabs.harvard.edu/abs/2021sf2a.conf..129M}{\color{blue}129}

\bibitem[{{Maire} {et~al.}(2016){Maire}, {Langlois}, {Dohlen}, {Lagrange},
  {Gratton}, {Chauvin}, {Desidera}, {Girard}, {Milli}, {Vigan}, {Zins},
  {Delorme}, {Beuzit}, {Claudi}, {Feldt}, {Mouillet}, {Puget}, {Turatto}, \&
  {Wildi}}]{maire16}
{Maire}, A.-L., {Langlois}, M., {Dohlen}, K., {et~al.} 2016,
  \href{http://dx.doi.org/10.1117/12.2233013}{\color{magenta}Proc.~SPIE},
  \href{https://ui.adsabs.harvard.edu/abs/2016SPIE.9908E..34M}{\color{blue}9908},
  \href{https://ui.adsabs.harvard.edu/abs/2016SPIE.9908E..34M}{\color{blue}990834}

\bibitem[{{Mesa} {et~al.}(2023){Mesa}, {Gratton}, {Kervella}, {Bonavita},
  {Desidera}, {D'Orazi}, {Marino}, {Zurlo}, \& {Rigliaco}}]{mesa23}
{Mesa}, D., {Gratton}, R., {Kervella}, P., {et~al.} 2023,
  \href{http://dx.doi.org/10.1051/0004-6361/202345865}{\color{magenta}\aap},
  \href{https://ui.adsabs.harvard.edu/abs/2023A&A...672A..93M}{\color{blue}672},
  \href{https://ui.adsabs.harvard.edu/abs/2023A&A...672A..93M}{\color{blue}A93}

\bibitem[{{Monnier} {et~al.}(2019){Monnier}, {Harries}, {Bae}, {Setterholm},
  {Laws}, {Aarnio}, {Adams}, {Andrews}, {Calvet}, {Espaillat}, {Hartmann},
  {Kraus}, {McClure}, {Miller}, {Oppenheimer}, {Wilner}, \& {Zhu}}]{monnier19}
{Monnier}, J.~D., {Harries}, T.~J., {Bae}, J., {et~al.} 2019,
  \href{http://dx.doi.org/10.3847/1538-4357/aafe87}{\color{magenta}\apj},
  \href{https://ui.adsabs.harvard.edu/abs/2019ApJ...872..122M}{\color{blue}872},
  \href{https://ui.adsabs.harvard.edu/abs/2019ApJ...872..122M}{\color{blue}122}

\bibitem[{{Montesinos} \& {Cuello}(2018)}]{montesinos18}
{Montesinos}, M., \& {Cuello}, N. 2018,
  \href{http://dx.doi.org/10.1093/mnrasl/sly001}{\color{magenta}\mnras},
  \href{https://ui.adsabs.harvard.edu/abs/2018MNRAS.475L..35M}{\color{blue}475},
  \href{https://ui.adsabs.harvard.edu/abs/2018MNRAS.475L..35M}{\color{blue}L35}

\bibitem[{{Nielsen} {et~al.}(2020){Nielsen}, {De Rosa}, {Wang}, {Sahlmann},
  {Kalas}, {Duch{\^e}ne}, {Rameau}, {Marley}, {Saumon}, {Macintosh},
  {Millar-Blanchaer}, {Nguyen}, {Ammons}, {Bailey}, {Barman}, {Bulger},
  {Chilcote}, {Cotten}, {Doyon}, {Esposito}, {Fitzgerald}, {Follette},
  {Gerard}, {Goodsell}, {Graham}, {Greenbaum}, {Hibon}, {Hung}, {Ingraham},
  {Konopacky}, {Larkin}, {Maire}, {Marchis}, {Marois}, {Metchev},
  {Oppenheimer}, {Palmer}, {Patience}, {Perrin}, {Poyneer}, {Pueyo}, {Rajan},
  {Rantakyr{\"o}}, {Ruffio}, {Savransky}, {Schneider}, {Sivaramakrishnan},
  {Song}, {Soummer}, {Thomas}, {Wallace}, {Ward-Duong}, {Wiktorowicz}, \&
  {Wolff}}]{nielsen20}
{Nielsen}, E.~L., {De Rosa}, R.~J., {Wang}, J.~J., {et~al.} 2020,
  \href{http://dx.doi.org/10.3847/1538-3881/ab5b92}{\color{magenta}\aj},
  \href{https://ui.adsabs.harvard.edu/abs/2020AJ....159...71N}{\color{blue}159},
  \href{https://ui.adsabs.harvard.edu/abs/2020AJ....159...71N}{\color{blue}71}

\bibitem[{{Nowak} {et~al.}(2020){Nowak}, {Lacour}, {Lagrange}, {Rubini},
  {Wang}, {Stolker}, {Abuter}, {Amorim}, {Asensio-Torres}, {Baub{\"o}ck},
  {Benisty}, {Berger}, {Beust}, {Blunt}, {Boccaletti}, {Bonnefoy}, {Bonnet},
  {Brandner}, {Cantalloube}, {Charnay}, {Choquet}, {Christiaens}, {Cl{\'e}net},
  {Coud{\'e} Du Foresto}, {Cridland}, {de Zeeuw}, {Dembet}, {Dexter},
  {Drescher}, {Duvert}, {Eckart}, {Eisenhauer}, {Gao}, {Garcia}, {Garcia
  Lopez}, {Gardner}, {Gendron}, {Genzel}, {Gillessen}, {Girard}, {Grandjean},
  {Haubois}, {Hei{\ss}el}, {Henning}, {Hinkley}, {Hippler}, {Horrobin},
  {Houll{\'e}}, {Hubert}, {Jim{\'e}nez-Rosales}, {Jocou}, {Kammerer},
  {Kervella}, {Keppler}, {Kreidberg}, {Kulikauskas}, {Lapeyr{\`e}re}, {Le
  Bouquin}, {L{\'e}na}, {M{\'e}rand}, {Maire}, {Molli{\`e}re}, {Monnier},
  {Mouillet}, {M{\"u}ller}, {Nasedkin}, {Ott}, {Otten}, {Paumard}, {Paladini},
  {Perraut}, {Perrin}, {Pueyo}, {Pfuhl}, {Rameau}, {Rodet},
  {Rodr{\'\i}guez-Coira}, {Rousset}, {Scheithauer}, {Shangguan}, {Stadler},
  {Straub}, {Straubmeier}, {Sturm}, {Tacconi}, {van Dishoeck}, {Vigan},
  {Vincent}, {von Fellenberg}, {Ward-Duong}, {Widmann}, {Wieprecht},
  {Wiezorrek}, {Woillez}, \& {GRAVITY Collaboration}}]{nowak20}
{Nowak}, M., {Lacour}, S., {Lagrange}, A.~M., {et~al.} 2020,
  \href{http://dx.doi.org/10.1051/0004-6361/202039039}{\color{magenta}\aap},
  \href{https://ui.adsabs.harvard.edu/abs/2020A&A...642L...2N}{\color{blue}642},
  \href{https://ui.adsabs.harvard.edu/abs/2020A&A...642L...2N}{\color{blue}L2}

\bibitem[{{Ohta} {et~al.}(2016){Ohta}, {Fukagawa}, {Sitko}, {Muto}, {Kraus},
  {Grady}, {Wisniewski}, {Swearingen}, {Shibai}, {Sumi}, {Hashimoto}, {Kudo},
  {Kusakabe}, {Momose}, {Okamoto}, {Kotani}, {Takami}, {Currie}, {Thalmann},
  {Janson}, {Akiyama}, {Follette}, {Mayama}, {Abe}, {Brandner}, {Brandt},
  {Carson}, {Egner}, {Feldt}, {Goto}, {Guyon}, {Hayano}, {Hayashi}, {Hayashi},
  {Henning}, {Hodapp}, {Ishii}, {Iye}, {Kandori}, {Knapp}, {Kuzuhara}, {Kwon},
  {Matsuo}, {McElwain}, {Miyama}, {Morino}, {Moro-Mart{\'\i}n}, {Nishimura},
  {Pyo}, {Serabyn}, {Suenaga}, {Suto}, {Suzuki}, {Takahashi}, {Takami},
  {Takato}, {Terada}, {Tomono}, {Turner}, {Usuda}, {Watanabe}, {Yamada}, \&
  {Tamura}}]{ohta16}
{Ohta}, Y., {Fukagawa}, M., {Sitko}, M.~L., {et~al.} 2016,
  \href{http://dx.doi.org/10.1093/pasj/psw051}{\color{magenta}\pasj},
  \href{https://ui.adsabs.harvard.edu/abs/2016PASJ...68...53O}{\color{blue}68},
  \href{https://ui.adsabs.harvard.edu/abs/2016PASJ...68...53O}{\color{blue}53}

\bibitem[{{Ren} {et~al.}(2018){Ren}, {Dong}, {Esposito}, {Pueyo}, {Debes},
  {Poteet}, {Choquet}, {Benisty}, {Chiang}, {Grady}, {Hines}, {Schneider}, \&
  {Soummer}}]{ren18}
{Ren}, B., {Dong}, R., {Esposito}, T.~M., {et~al.} 2018,
  \href{http://dx.doi.org/10.3847/2041-8213/aab7f5}{\color{magenta}\apjl},
  \href{https://ui.adsabs.harvard.edu/abs/2018ApJ...857L...9R}{\color{blue}857},
  \href{https://ui.adsabs.harvard.edu/abs/2018ApJ...857L...9R}{\color{blue}L9}

\bibitem[{{Ren} {et~al.}(2020){Ren}, {Dong}, {van Holstein}, {Ruffio},
  {Calvin}, {Girard}, {Benisty}, {Boccaletti}, {Esposito}, {Choquet}, {Mawet},
  {Pueyo}, {Stolker}, {Chiang}, {Boer}, {Debes}, {Garufi}, {Grady}, {Hines},
  {Maire}, {M{\'e}nard}, {Millar-Blanchaer}, {Perrin}, {Poteet}, \&
  {Schneider}}]{ren20}
{Ren}, B., {Dong}, R., {van Holstein}, R.~G., {et~al.} 2020,
  \href{http://dx.doi.org/10.3847/2041-8213/aba43e}{\color{magenta}\apjl},
  \href{https://ui.adsabs.harvard.edu/abs/2020ApJ...898L..38R}{\color{blue}898},
  \href{https://ui.adsabs.harvard.edu/abs/2020ApJ...898L..38R}{\color{blue}L38}

\bibitem[{{Ren} {et~al.}(2023){Ren}, {Benisty}, {Ginski}, {Tazaki}, {Wallack},
  {Milli}, {Garufi}, {Bae}, {Facchini}, {M{\'e}nard}, {Pinilla}, {Swastik},
  {Teague}, \& {Wahhaj}}]{ren23}
{Ren}, B.~B., {Benisty}, M., {Ginski}, C., {et~al.} 2023,
  \href{https://arxiv.org/abs/2310.08589}{\color{magenta}arXiv},
  \href{https://ui.adsabs.harvard.edu/abs/2023arXiv231008589R}{\color{blue}arXiv:2310.08589}

\bibitem[{{Safonov} {et~al.}(2022){Safonov}, {Strakhov}, {Goliguzova}, \&
  {Voziakova}}]{safonov22}
{Safonov}, B.~S., {Strakhov}, I.~A., {Goliguzova}, M.~V., \& {Voziakova}, O.~V.
  2022, \href{http://dx.doi.org/10.3847/1538-3881/ac36cb}{\color{magenta}\aj},
  \href{https://ui.adsabs.harvard.edu/abs/2022AJ....163...31S}{\color{blue}163},
  \href{https://ui.adsabs.harvard.edu/abs/2022AJ....163...31S}{\color{blue}31}

\bibitem[{{Schmid} {et~al.}(2018){Schmid}, {Bazzon}, {Roelfsema}, {Mouillet},
  {Milli}, {Menard}, {Gisler}, {Hunziker}, {Pragt}, {Dominik}, {Boccaletti},
  {Ginski}, {Abe}, {Antoniucci}, {Avenhaus}, {Baruffolo}, {Baudoz}, {Beuzit},
  {Carbillet}, {Chauvin}, {Claudi}, {Costille}, {Daban}, {de Haan}, {Desidera},
  {Dohlen}, {Downing}, {Elswijk}, {Engler}, {Feldt}, {Fusco}, {Girard},
  {Gratton}, {Hanenburg}, {Henning}, {Hubin}, {Joos}, {Kasper}, {Keller},
  {Langlois}, {Lagadec}, {Martinez}, {Mulder}, {Pavlov}, {Podio}, {Puget},
  {Quanz}, {Rigal}, {Salasnich}, {Sauvage}, {Schuil}, {Siebenmorgen}, {Sissa},
  {Snik}, {Suarez}, {Thalmann}, {Turatto}, {Udry}, {van Duin}, {van Holstein},
  {Vigan}, \& {Wildi}}]{schmid18}
{Schmid}, H.~M., {Bazzon}, A., {Roelfsema}, R., {et~al.} 2018,
  \href{http://dx.doi.org/10.1051/0004-6361/201833620}{\color{magenta}\aap},
  \href{https://ui.adsabs.harvard.edu/abs/2018A&A...619A...9S}{\color{blue}619},
  \href{https://ui.adsabs.harvard.edu/abs/2018A&A...619A...9S}{\color{blue}A9}

\bibitem[{{Shuai} {et~al.}(2022){Shuai}, {Ren}, {Dong}, {Zhou}, {Pueyo}, {De
  Rosa}, {Fang}, \& {Mawet}}]{shuai22}
{Shuai}, L., {Ren}, B.~B., {Dong}, R., {et~al.} 2022,
  \href{http://dx.doi.org/10.3847/1538-4365/ac98fd}{\color{magenta}\apjs},
  \href{https://ui.adsabs.harvard.edu/abs/2022ApJS..263...31S}{\color{blue}263},
  \href{https://ui.adsabs.harvard.edu/abs/2022ApJS..263...31S}{\color{blue}31}

\bibitem[{{Spiegel} \& {Burrows}(2012)}]{spiegel12}
{Spiegel}, D.~S., \& {Burrows}, A. 2012,
  \href{http://dx.doi.org/10.1088/0004-637X/745/2/174}{\color{magenta}\apj},
  \href{https://ui.adsabs.harvard.edu/abs/2012ApJ...745..174S}{\color{blue}745},
  \href{https://ui.adsabs.harvard.edu/abs/2012ApJ...745..174S}{\color{blue}174}

\bibitem[{{Stolker} {et~al.}(2019){Stolker}, {Bonse}, {Quanz}, {Amara},
  {Cugno}, {Bohn}, \& {Boehle}}]{pynpoint}
{Stolker}, T., {Bonse}, M.~J., {Quanz}, S.~P., {et~al.} 2019,
  \href{http://dx.doi.org/10.1051/0004-6361/201834136}{\color{magenta}\aap},
  \href{https://ui.adsabs.harvard.edu/abs/2019A&A...621A..59S}{\color{blue}621},
  \href{https://ui.adsabs.harvard.edu/abs/2019A&A...621A..59S}{\color{blue}A59}

\bibitem[{{van Holstein} {et~al.}(2017){van Holstein}, {Snik}, {Girard}, {de
  Boer}, {Ginski}, {Keller}, {Stam}, {Beuzit}, {Mouillet}, {Kasper},
  {Langlois}, {Zurlo}, {de Kok}, \& {Vigan}}]{irdap1}
{van Holstein}, R.~G., {Snik}, F., {Girard}, J.~H., {et~al.} 2017,
  \href{http://dx.doi.org/10.1117/12.2272554}{\color{magenta}Proc.~SPIE},
  \href{https://ui.adsabs.harvard.edu/abs/2017SPIE10400E..15V}{\color{blue}10400},
  \href{https://ui.adsabs.harvard.edu/abs/2017SPIE10400E..15V}{\color{blue}1040015}

\bibitem[{{van Holstein} {et~al.}(2020){van Holstein}, {Girard}, {de Boer},
  {Snik}, {Milli}, {Stam}, {Ginski}, {Mouillet}, {Wahhaj}, {Schmid}, {Keller},
  {Langlois}, {Dohlen}, {Vigan}, {Pohl}, {Carbillet}, {Fantinel}, {Maurel},
  {Orign{\'e}}, {Petit}, {Ramos}, {Rigal}, {Sevin}, {Boccaletti}, {Le
  Coroller}, {Dominik}, {Henning}, {Lagadec}, {M{\'e}nard}, {Turatto}, {Udry},
  {Chauvin}, {Feldt}, \& {Beuzit}}]{irdap2}
{van Holstein}, R.~G., {Girard}, J.~H., {de Boer}, J., {et~al.} 2020,
  \href{http://dx.doi.org/10.1051/0004-6361/201834996}{\color{magenta}\aap},
  \href{https://ui.adsabs.harvard.edu/abs/2020A&A...633A..64V}{\color{blue}633},
  \href{https://ui.adsabs.harvard.edu/abs/2020A&A...633A..64V}{\color{blue}A64}

\bibitem[{{Vieira} {et~al.}(2003){Vieira}, {Corradi}, {Alencar}, {Mendes},
  {Torres}, {Quast}, {Guimar{\~a}es}, \& {da Silva}}]{vieira03}
{Vieira}, S.~L.~A., {Corradi}, W.~J.~B., {Alencar}, S.~H.~P., {et~al.} 2003,
  \href{http://dx.doi.org/10.1086/379553}{\color{magenta}\aj},
  \href{https://ui.adsabs.harvard.edu/abs/2003AJ....126.2971V}{\color{blue}126},
  \href{https://ui.adsabs.harvard.edu/abs/2003AJ....126.2971V}{\color{blue}2971}

\bibitem[{{Vigan} {et~al.}(2021){Vigan}, {Fontanive}, {Meyer}, {Biller},
  {Bonavita}, {Feldt}, {Desidera}, {Marleau}, {Emsenhuber}, {Galicher}, {Rice},
  {Forgan}, {Mordasini}, {Gratton}, {Le Coroller}, {Maire}, {Cantalloube},
  {Chauvin}, {Cheetham}, {Hagelberg}, {Lagrange}, {Langlois}, {Bonnefoy},
  {Beuzit}, {Boccaletti}, {D'Orazi}, {Delorme}, {Dominik}, {Henning}, {Janson},
  {Lagadec}, {Lazzoni}, {Ligi}, {Menard}, {Mesa}, {Messina}, {Moutou},
  {M{\"u}ller}, {Perrot}, {Samland}, {Schmid}, {Schmidt}, {Sissa}, {Turatto},
  {Udry}, {Zurlo}, {Abe}, {Antichi}, {Asensio-Torres}, {Baruffolo}, {Baudoz},
  {Baudrand}, {Bazzon}, {Blanchard}, {Bohn}, {Brown Sevilla}, {Carbillet},
  {Carle}, {Cascone}, {Charton}, {Claudi}, {Costille}, {De Caprio},
  {Delboulb{\'e}}, {Dohlen}, {Engler}, {Fantinel}, {Feautrier}, {Fusco},
  {Gigan}, {Girard}, {Giro}, {Gisler}, {Gluck}, {Gry}, {Hubin}, {Hugot},
  {Jaquet}, {Kasper}, {Le Mignant}, {Llored}, {Madec}, {Magnard}, {Martinez},
  {Maurel}, {M{\"o}ller-Nilsson}, {Mouillet}, {Moulin}, {Orign{\'e}}, {Pavlov},
  {Perret}, {Petit}, {Pragt}, {Puget}, {Rabou}, {Ramos}, {Rickman}, {Rigal},
  {Rochat}, {Roelfsema}, {Rousset}, {Roux}, {Salasnich}, {Sauvage}, {Sevin},
  {Soenke}, {Stadler}, {Suarez}, {Wahhaj}, {Weber}, \& {Wildi}}]{vigan21}
{Vigan}, A., {Fontanive}, C., {Meyer}, M., {et~al.} 2021,
  \href{http://dx.doi.org/10.1051/0004-6361/202038107}{\color{magenta}\aap},
  \href{https://ui.adsabs.harvard.edu/abs/2021A&A...651A..72V}{\color{blue}651},
  \href{https://ui.adsabs.harvard.edu/abs/2021A&A...651A..72V}{\color{blue}A72}

\bibitem[{{Wagner} {et~al.}(2019){Wagner}, {Stone}, {Spalding}, {Apai}, {Dong},
  {Ertel}, {Leisenring}, \& {Webster}}]{wagner19}
{Wagner}, K., {Stone}, J.~M., {Spalding}, E., {et~al.} 2019,
  \href{http://dx.doi.org/10.3847/1538-4357/ab32ea}{\color{magenta}\apj},
  \href{https://ui.adsabs.harvard.edu/abs/2019ApJ...882...20W}{\color{blue}882},
  \href{https://ui.adsabs.harvard.edu/abs/2019ApJ...882...20W}{\color{blue}20}

\bibitem[{{Wagner} {et~al.}(2023){Wagner}, {Stone}, {Skemer}, {Ertel}, {Dong},
  {Apai}, {Spalding}, {Leisenring}, {Sitko}, {Kratter}, {Barman}, {Marley},
  {Miles}, {Boccaletti}, {Assani}, {Bayyari}, {Uyama}, {Woodward}, {Hinz},
  {Briesemeister}, {Lawson}, {M{\'e}nard}, {Pantin}, {Russell}, {Skrutskie}, \&
  {Wisniewski}}]{wagner23}
{Wagner}, K., {Stone}, J., {Skemer}, A., {et~al.} 2023,
  \href{http://dx.doi.org/10.1038/s41550-023-02028-3}{\color{magenta}NatAs},
  \href{https://ui.adsabs.harvard.edu/abs/2023NatAs.tmp..146W}{\color{blue}7},
  \href{https://ui.adsabs.harvard.edu/abs/2023NatAs.tmp..146W}{\color{blue}1208}

\bibitem[{{Wallack} {et~al.}(2023){Wallack}, {Ruffio}, {Ruane},
  {et~al.}}]{wallack23}
{Wallack}, N.~L., {Ruffio}, J.-B., {Ruane}, G., {et~al.} 2023, \aj, in press

\bibitem[{{Wang} {et~al.}(2020){Wang}, {Ginzburg}, {Ren}, {Wallack}, {Gao},
  {Mawet}, {Bond}, {Cetre}, {Wizinowich}, {De Rosa}, {Ruane}, {Liu}, {Absil},
  {Alvarez}, {Baranec}, {Choquet}, {Chun}, {Defr{\`e}re}, {Delorme},
  {Duch{\^e}ne}, {Forsberg}, {Ghez}, {Guyon}, {Hall}, {Huby}, {Jolivet},
  {Jensen-Clem}, {Jovanovic}, {Karlsson}, {Lilley}, {Matthews}, {M{\'e}nard},
  {Meshkat}, {Millar-Blanchaer}, {Ngo}, {Orban de Xivry}, {Pinte}, {Ragland},
  {Serabyn}, {Catal{\'a}n}, {Wang}, {Wetherell}, {Williams}, {Ygouf}, \&
  {Zuckerman}}]{wang20}
{Wang}, J.~J., {Ginzburg}, S., {Ren}, B., {et~al.} 2020,
  \href{http://dx.doi.org/10.3847/1538-3881/ab8aef}{\color{magenta}\aj},
  \href{https://ui.adsabs.harvard.edu/abs/2020AJ....159..263W}{\color{blue}159},
  \href{https://ui.adsabs.harvard.edu/abs/2020AJ....159..263W}{\color{blue}263}

\bibitem[{{Willson} {et~al.}(2019){Willson}, {Kraus}, {Kluska}, {Monnier},
  {Cure}, {Sitko}, {Aarnio}, {Ireland}, {Rizzuto}, {Hone}, {Kreplin},
  {Andrews}, {Calvet}, {Espaillat}, {Fukagawa}, {Harries}, {Hinkley}, {Kanaan},
  {Muto}, \& {Wilner}}]{willson19}
{Willson}, M., {Kraus}, S., {Kluska}, J., {et~al.} 2019,
  \href{http://dx.doi.org/10.1051/0004-6361/201630215}{\color{magenta}\aap},
  \href{https://ui.adsabs.harvard.edu/abs/2019A&A...621A...7W}{\color{blue}621},
  \href{https://ui.adsabs.harvard.edu/abs/2019A&A...621A...7W}{\color{blue}A7}

\bibitem[{{W{\"o}lfer} {et~al.}(2023){W{\"o}lfer}, {Facchini}, {van der Marel},
  {van Dishoeck}, {Benisty}, {Bohn}, {Francis}, {Izquierdo}, \&
  {Teague}}]{woelfer23}
{W{\"o}lfer}, L., {Facchini}, S., {van der Marel}, N., {et~al.} 2023,
  \href{http://dx.doi.org/10.1051/0004-6361/202243601}{\color{magenta}\aap},
  \href{https://ui.adsabs.harvard.edu/abs/2023A&A...670A.154W}{\color{blue}670},
  \href{https://ui.adsabs.harvard.edu/abs/2023A&A...670A.154W}{\color{blue}A154}

\bibitem[{{Xie} {et~al.}(2021){Xie}, {Ren}, {Dong}, {Pueyo}, {Ruffio}, {Fang},
  {Mawet}, \& {Stolker}}]{xie21}
{Xie}, C., {Ren}, B., {Dong}, R., {et~al.} 2021,
  \href{http://dx.doi.org/10.3847/2041-8213/abd241}{\color{magenta}\apjl},
  \href{https://ui.adsabs.harvard.edu/abs/2021ApJ...906L...9X}{\color{blue}906},
  \href{https://ui.adsabs.harvard.edu/abs/2021ApJ...906L...9X}{\color{blue}L9}

\bibitem[{{Xie} {et~al.}(2023){Xie}, {Ren}, {Dong}, {Choquet}, {Vigan},
  {Gonzalez}, {Wagner}, {Fang}, \& {Ubeira-Gabellini}}]{xie23}
{Xie}, C., {Ren}, B.~B., {Dong}, R., {et~al.} 2023,
  \href{http://dx.doi.org/10.1051/0004-6361/202346305}{\color{magenta}\aap},
  \href{https://ui.adsabs.harvard.edu/abs/2023A&A...675L...1X}{\color{blue}675},
  \href{https://ui.adsabs.harvard.edu/abs/2023A&A...675L...1X}{\color{blue}L1}

\bibitem[{{Yu} {et~al.}(2019){Yu}, {Ho}, \& {Zhu}}]{yu19}
{Yu}, S.-Y., {Ho}, L.~C., \& {Zhu}, Z. 2019,
  \href{http://dx.doi.org/10.3847/1538-4357/ab1d65}{\color{magenta}\apj},
  \href{https://ui.adsabs.harvard.edu/abs/2019ApJ...877..100Y}{\color{blue}877},
  \href{https://ui.adsabs.harvard.edu/abs/2019ApJ...877..100Y}{\color{blue}100}

\bibitem[{{Zhang} {et~al.}(2018){Zhang}, {Zhu}, {Huang}, {Guzm{\'a}n},
  {Andrews}, {Birnstiel}, {Dullemond}, {Carpenter}, {Isella}, {P{\'e}rez},
  {Benisty}, {Wilner}, {Baruteau}, {Bai}, \& {Ricci}}]{zhang18}
{Zhang}, S., {Zhu}, Z., {Huang}, J., {et~al.} 2018,
  \href{http://dx.doi.org/10.3847/2041-8213/aaf744}{\color{magenta}\apjl},
  \href{https://ui.adsabs.harvard.edu/abs/2018ApJ...869L..47Z}{\color{blue}869},
  \href{https://ui.adsabs.harvard.edu/abs/2018ApJ...869L..47Z}{\color{blue}L47}

\bibitem[{{Zurlo} {et~al.}(2014){Zurlo}, {Vigan}, {Mesa}, {Gratton}, {Moutou},
  {Langlois}, {Claudi}, {Pueyo}, {Boccaletti}, {Baruffolo}, {Beuzit},
  {Costille}, {Desidera}, {Dohlen}, {Feldt}, {Fusco}, {Henning}, {Kasper},
  {Martinez}, {Moeller-Nilsson}, {Mouillet}, {Pavlov}, {Puget}, {Sauvage},
  {Turatto}, {Udry}, {Vakili}, {Waters}, \& {Wildi}}]{zurlo14}
{Zurlo}, A., {Vigan}, A., {Mesa}, D., {et~al.} 2014,
  \href{http://dx.doi.org/10.1051/0004-6361/201424204}{\color{magenta}\aap},
  \href{https://ui.adsabs.harvard.edu/abs/2014A&A...572A..85Z}{\color{blue}572},
  \href{https://ui.adsabs.harvard.edu/abs/2014A&A...572A..85Z}{\color{blue}A85}

\end{thebibliography}

\appendix
\section{Alternative Measurements}\label{sec-app}
We present the measured values for the entirety of the S1 and S2 arms here. Given the ${\sim}30^\circ$ inclination effects \citep{dong16spiralprojection} and shadowing effects \citep{montesinos18} in Fig.~\ref{fig1}, the motion of entire spirals is biased. However, we present the results here to show the global motion, which might be indicative of eccentric drivers \citep{calcino20, xie23}.

\begin{figure}[htb!]
	\centering
	\includegraphics[width=0.51\textwidth]{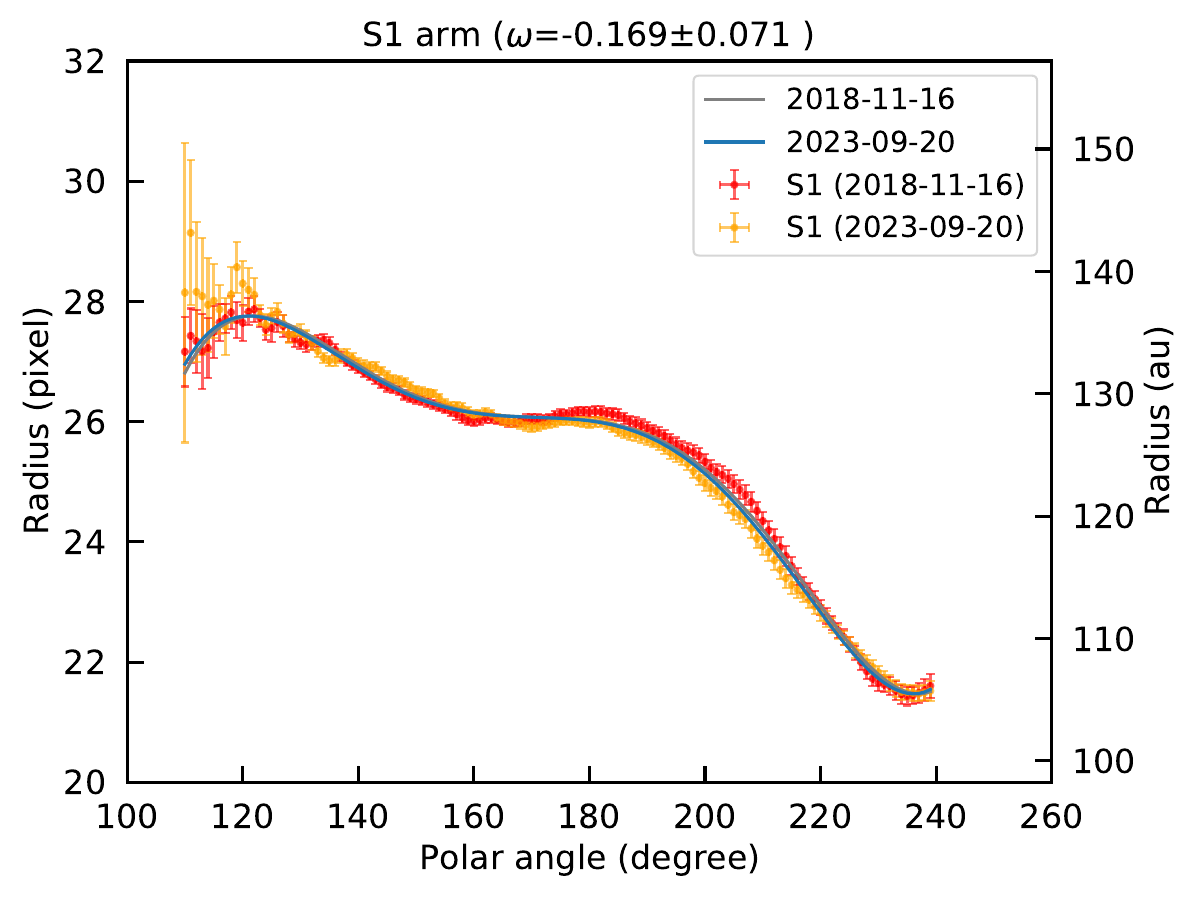}
    \caption{Peak locations of spiral arm S1 (full arm) in polar coordinates after correction for viewing geometry. The solid curves represent the best-fit model spiral for the peak locations (dot points) between 2018 and 2023 ($t=4.84$~yr), assuming the companion-driven scenario. The derived angular velocities of spiral pattern motion are $-0\fdg17\pm0\fdg07$~yr$^{-1}$.}
    \label{fig-a2}
\end{figure}

\begin{figure}[htb!]
	\centering
	\includegraphics[width=0.51\textwidth]{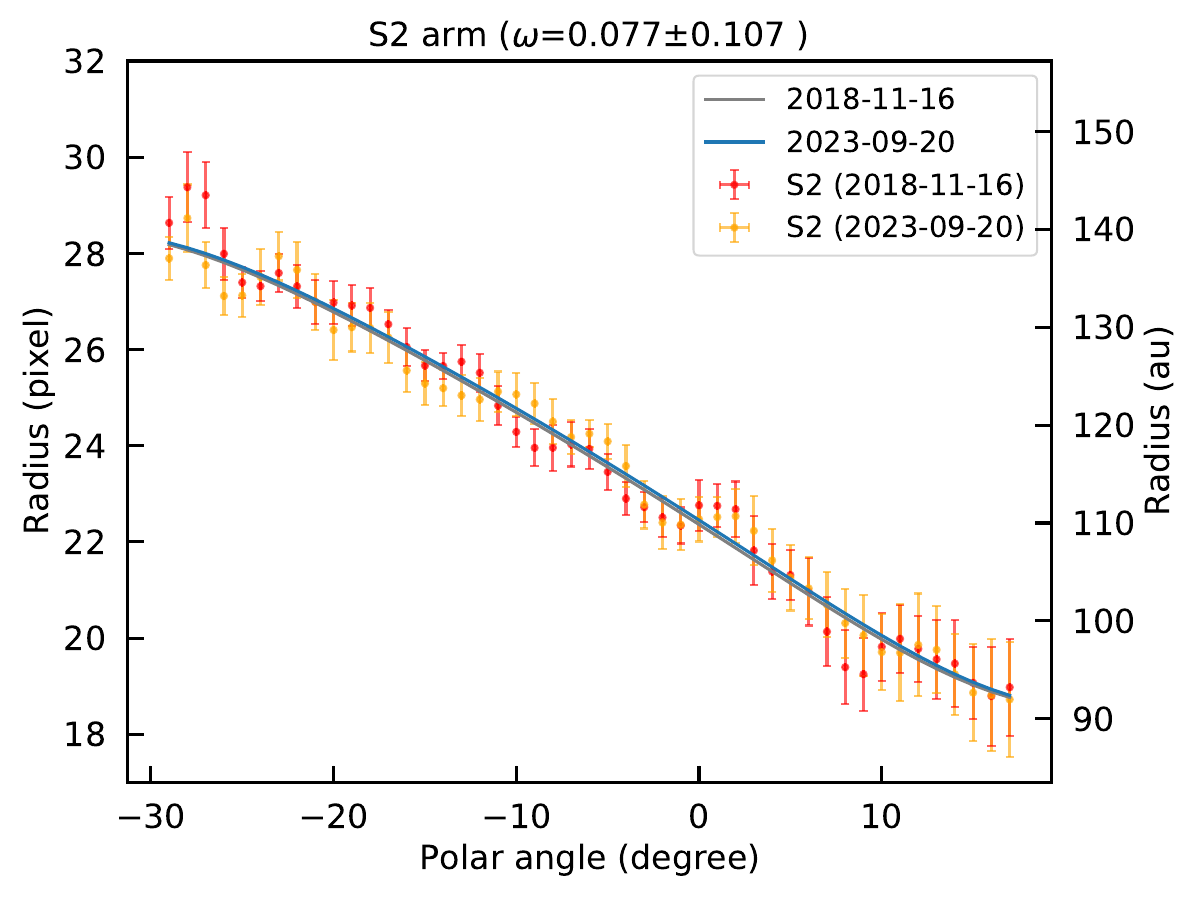}
    \caption{Peak locations of spiral arm S2 in polar coordinates after correction for viewing geometry. The solid curves represent the best-fit model spiral for the peak locations (dot points) between 2018 and 2023 ($t=4.84$~yr), assuming the companion-driven scenario. The derived angular velocities of spiral pattern motion are $0\fdg08\pm0\fdg10$~yr$^{-1}$.}
    \label{fig-a3}
\end{figure}


\end{CJK*}
\end{document}